\documentclass[aps,prl,reprint,superscriptaddress]{revtex4-2}
\usepackage{graphicx}
\usepackage[dvipsnames]{xcolor}
\usepackage{multirow}
\usepackage{amsmath}
\usepackage[normalem]{ulem}

\begin{document}

\title{Simple production of cellulose nanofibril microcapsules and the rheology of their suspensions}

\author{Abhishek P. Dhand}
\affiliation{Department of Chemical and Biomolecular Engineering, University of Pennsylvania, Philadelphia, Pennsylvania 19104}

\author{Ryan Poling-Skutvik}
\email{ryanps@uri.edu}
\affiliation{Department of Chemical Engineering, University of Rhode Island, Kingston, RI 02881}

\author{Chinedum O. Osuji}
\email{cosuji@seas.upenn.edu}
\affiliation{Department of Chemical and Biomolecular Engineering, University of Pennsylvania, Philadelphia, Pennsylvania 19104}

\date{\today}

\begin{abstract}
Microcapsules are commonly used in applications ranging from therapeutics to personal care products due to their ability to deliver encapsulated species through their porous shells. Here, we demonstrate a simple and scalable approach to fabricate microcapsules with porous shells by interfacial complexation of cellulose nanofibrils and oleylamine, and investigate the rheological properties of suspensions of the resulting microcapsules. The suspensions of neat capsules are viscous liquids whose viscosity increases with volume fraction according to a modified Kreiger-Dougherty relation with a maximum packing fraction of 0.73 and an intrinsic viscosity of 4. When polyacrylic acid (PAA) is added to the internal phase of the microcapsule, however, the suspensions become elastic and display yield stresses with power-law dependencies on capsule volume fraction and PAA concentration. The elasticity appears to originate from associative interactions between microcapsules induced by PAA that resides within the microcapsule shells. These results demonstrate that it is possible to tune the rheological properties of microcapsule suspensions by changing only the composition of the internal phase, thereby providing a novel method to tailor complex fluid rheology.
\end{abstract}

\maketitle

\section{Introduction}
Microcapsules are droplets of liquid encapsulated within a thin, solid shell.\cite{Yow2006,Lensen2008,Amstad2017} Whereas two immiscible liquids can be chemically stabilized with a surfactant to form an emulsion, the internal and external phases of a microcapsule are \emph{physically} separated by the shell. The majority of capsule syntheses use microfluidic\cite{Zhang2012, Kaufman2014, Kaufman2017} or bulk\cite{DupreDeBaubigny2017,Song2019} approaches to drive the complexation or precipitation of species such as graphene,\cite{Liu2020} nanoparticles,\cite{Dinsmore2002} or polymers,\cite{Discher1999} at the droplet interface. The resulting shells are commonly semipermeable to allow for exchange between the encapsulated phase and the surrounding environment. This exchange mechanism makes microcapsules ideally suited for use in drug delivery and controlled release applications,\cite{Bedard2010} self-healing materials,\cite{Blaiszik2010} and cell and tissue cultures.\cite{Morimoto2009,Niepa2016,Manimaran2020} In each of these applications, suspensions of porous microcapsules must be processed and delivered to areas of interest before the exchange begins. Thus, it is critical to understand the rheology and behavior of microcapsule suspensions under shear. 

For suspensions of non-interacting hard spheres, the rheology depends completely on a single variable -- the volume fraction $\phi$.\cite{Mewis2012} At low $\phi$, hard sphere suspensions are Newtonian with a viscosity that increases as $\eta = \eta_\mathrm{s}(1+[\eta]\phi)$, where $[\eta] = 2.5$ is the intrinsic viscosity and $\eta_\mathrm{s}$ is the solvent viscosity. At larger $\phi$, the suspensions become non-Newtonian and shear thin with a zero-shear viscosity that follows the Krieger-Dougherty expression\cite{Wildemuth1984} $\eta_0 = \eta_\mathrm{s}(1-\phi/\phi_m)^{-[\eta]\phi_m}$, where $\phi_m \approx 0.64$ is the volume fraction at which the viscosity diverges. Beyond $\phi_m$, hard-sphere suspensions become glassy and no longer flow.\cite{Hunter2012} Microcapsules are deformable particles, however, whose interaction profiles are much softer than those of hard sphere particles.\cite{Liu1996} This softness or deformability becomes a second variable, in addition to $\phi$, controlling suspension viscosity.\cite{Vlassopoulos2014} The viscosity of soft sphere suspensions increases more gradually and diverges at higher volume fractions than that of hard sphere suspensions.\cite{Mattsson2009} If the compression and deformability of soft particles is accounted for, their suspension rheology can be closely mapped onto hard sphere predictions.\cite{Pellet2016,Gnan2019}  Microcapsules are unique when compared to other classes of soft particles, such as polymer-grafted nanoparticles,\cite{Poling-Skutvik2019} charged colloids,\cite{Holmqvist2010} and cross-linked microgels,\cite{Senff1999} because they have a solid shell surrounding an incompressible fluid. The flexibility and stiffness of this shell has been shown to control the capsule deformation under shear\cite{Smith2006, Finken2006, Gross2014} and induce novel rheological phenomena such as negative intrinsic viscosity\cite{Gao2012} and stream migration.\cite{Danker2009, Abreu2014}

In this paper, we demonstrate a facile and scalable method to produce microcapsules approximately 25-200 $\mu$m in diameter through a complexation between cellulose nanofibrils and oleylamine. The resulting microcapsules are stable in organic and aqueous suspensions and are porous to allow diffusion of encapsulated macromolecules into the continuous phase. Suspensions of these microcapsules are strongly shear-thinning with zero-shear viscosities that deviate from hard sphere predictions as a function of volume fraction. Interpreted through a modified Krieger-Dougherty expression, the suspensions of neat capsules exhibit an intrinsic viscosity of $4.1 \pm 1.6$ and a maximum packing fraction of $0.73 \pm 0.22$, which may reflect the non-sphericity, polydispersity, and softness of the CNF capsules. We tailor the interactions between microcapsules in suspension by incorporating polymeric additives to the internal phase. These additives induce strong attractive interactions between the microcapsules, resulting in the development of a yield stress that scales with microcapsule volume fraction $\phi$ and additive concentration. We confirm that the modified microcapsules quiescently structure into an elastic network of clusters that break apart under shear, generating hysteresis in the suspension flow behavior. Our results demonstrate that additives to the \emph{internal} phase of microcapsules can control the solution structure and rheology.

\section{Materials and Methods}

\subsection{Millifluidic synthesis of microcapsules}
A stock suspension of TEMPO-modified cellulose nanofibrils (CNF) was acquired from the Process Development Center at the University of Maine at a concentration of \hbox{$1.1$ wt.\%} and a surface charge concentration of 1.5 mM per gram of dry CNF. The stock solution of CNF was diluted to a concentration of 0.3 wt\% to form the aqueous dispersed phase. The organic, continuous phase was formed by dissolving oleylamine (OA, Millipore Sigma) in toluene at 2.5 wt.\% composition. CNF/OA capsules were synthesized using a millifluidic junction device fabricated with 1/32-inch PTFE tubing connected to polypropylene T-junction device. The aqueous CNF suspension was flowed at 0.01 to 0.04 mL/min through the T-junction into an external, continuous phase of 2.5 wt.\% OA/Toluene solution with a flowrate of 0.2 to 0.4 mL/min. The length of the tubing was adjusted to allow for a residence time of at least 5 min to ensure complete capsule formation. The capsules were collected in a fresh 2.5 wt.\% OA/ Toluene solution. This millifluidic approach produced spherical capsules at a size of $580\pm200$ $\mu$m but at a low-throughput. To produce large enough volumes of capsules for rheological measurements, we transitioned to a batch synthesis approach.

\subsection{Batch synthesis of microcapsules}
A master batch of capsules was prepared by adding the aqueous CNF suspension to the OA/toluene solution at a volume fraction $\phi = 0.15$ and mixed using a high speed rotating mixer (3000 or 10,000 rpm, IKA T18 Ultra Turrax) for approximately two minutes until the aqueous phase was completely dispersed and the solution becomes turbid. The suspension was then stirred at 3000 rpm for five minutes to keep the microcapsules suspended and to allow for microcapsule shell formation. Microcapsule suspensions were concentrated from this master batch by decanting the supernatant. Suspensions of microcapsules modified with polyacrylic acid (PAA, approx. $M_\mathrm{w} = 250$ kDa, Millipore Sigma) were produced in a similar manner by first dissolving the desired concentration of PAA in the aqueous solution before mixing. Suspensions with depletant interactions were synthesized by dissolving polystyrene (PS, approx. $M_\mathrm{w} = 210$ kDa, Scientific Polymer Products, Inc.) to the continuous phase during density matching.

\subsection{Density matching}
 The density of the surrounding medium was matched to that of the microcapsules to avoid sedimentation during rheology measurements. Chloroform was added to toluene dropwise and until no sedimentation was observed under centrifugation at $1200\times$G. The final composition of the continuous phase was 64.5\% toluene by weight and the balance chloroform. Chloroform did not affect the stability of the microcapsules in suspension.
 
\subsection{Solvent transfer}
Optical microscopy experiments were conducted by transferring the microcapsules to an aqueous phase. Microcapsules suspended in toluene were centrifuged and the toluene decanted. The microcapsules were then washed with hexane, centrifuged, and the solvent decanted. Residual hexane was evaporated by gentle exposure to dry nitrogen gas. The remaining capsules were then washed twice with DI water before any measurements.

\subsection{Microscopy}
FITC-labelled Dextran (10 kDa, Millipore-Sigma) and rhodamine labeled PS microparticles (diameters of 0.8 $\mu$m and 2 $\mu$m, Thermo-Fisher Fluoro-Max) were suspended in the aqueous phase prior to fabrication to encapsulate them within the microcapsule shell. Microcapsules were characterized using bright field and fluorescence microscopy using a Zeiss Axiovert 200 M inverted microscope at magnifications of 5$\times$ and 20$\times$. The microcapsules were placed in a petri dish and immersed in either hexane or water and then imaged over time. 

\subsection{Rheology}
All rheological measurements were performed on TA ARES G2 rheometer using concentric cylinder geometry a bob diameter of 27.666 mm and a cup diameter of 30 mm. Oscillatory measurements were conducted at 25$^\circ$C and at an oscillation strain of $\gamma = 0.5$ \%. This strain was confirmed by an amplitude sweep to be in the linear viscoelastic regime for all microcapsule suspensions. Steady shear measurements were conducted with increasing and decreasing shear rates to assess sample stability and hysteresis. At each shear rate, the sample was sheared for 10 seconds to establish steady state and then averaged over an additional 5 seconds to quantify the stress. Sample evaporation was minimal in this geometry because of the small area exposed to the atmosphere relative to the volume of the sample. 

\section{Results and Discussion}

\subsection{Microcapsule synthesis and characterization}

\begin{figure*}[ht]
\includegraphics[width=\textwidth]{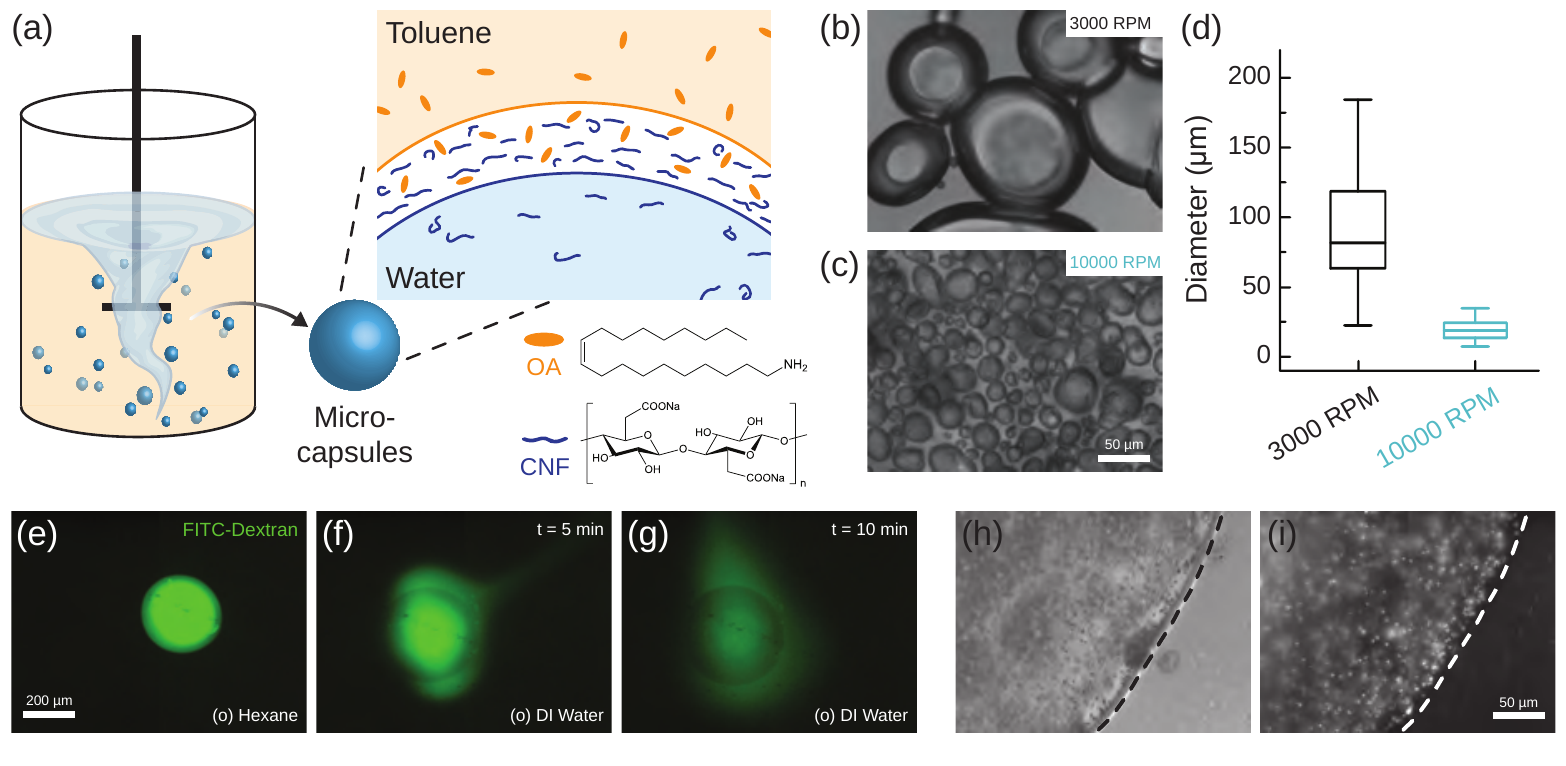}
\caption{(a) Schematic of microcapsule batch synthesis. (b,c) Micrographs  and (d) size distribution of microcapsules formed at different mixer speeds. Micrographs of FITC-dextran loaded microcapsules suspended in (e) hexane and in water after (f) $t = 5$ min and (g) $t = 10$ min. Microcapsules in water containing 2 $\mu$m microparticles under (h) brightfield and (i) fluorescent illumination. Dashed lines indicate microcapsule boundary.}
\label{fig:figure1}
\end{figure*}

We previously reported the synthesis of microcapsules formed by complexation of a random cationic copolymer (MADQUAT-co-BTA) with anionic cellulose nanofibrils (CNF).\cite{Kaufman2017} Here, we show that a similar complexation is achievable using oleylamine (OA), a commercially available small molecule. We first demonstrate the formation of microcapsules via the OA/CNF complexation using a millifluidic device. The CNF and OA complex at the droplet interface to form a thin, solid shell\cite{Calabrese2020a} that mechanically stabilizes the droplets to prevent coalescence. The resulting microcapsules are strong enough to survive solvent transfer out of the oil phase into an aqueous phase. When the microcapsules are suspended in an oil phase, the aqueous internal phase of the microcapsule remains trapped within the CNF shell (Fig.\ \ref{fig:figure1}). After solvent transfer into an aqueous phase, however, the microcapsules begin to exchange their internal phase with the surrounding solution through pores in the CNF shell. We visualize this exchange by encapsulating FITC-labelled dextran within the microcapsules and imaging the capsules over time. By contrast, larger objects such as 2 $\mu$m microparticles remain trapped within the porous shell and do not diffuse into the surrounding liquid. These measurements suggest that the pores in the microcapsule shell are on the order of 100 nm -- 1 $\mu$m. Confirming that we successfully synthesize porous microcapsules through the complexation of oleylamine and CNF, we now focus on characterizing the behavior of the capsules in suspension. 

Although a micro- or millifluidic synthesis produces uniform capsules with a controlled size, the process is difficult to scale to produce large volumes for bulk rheological characterization. As an alternative production method, we develop a simple bulk synthesis procedure in which the OA-rich oil and CNF suspension are combined in a high-shear mixer (Fig.\ \ref{fig:figure1}). The shear breaks the aqueous suspension into droplets and prevents coalescence as the OA and CNF complex, resulting in stable suspensions at a volume fraction $\phi = 0.15$. The size of the capsules is controlled by the speed of the mixer, with higher speeds generating smaller capsules (Fig.\ \ref{fig:figure1}). Suspensions are then concentrated by decanting excess solvent after centrifugation and subsequent resuspension. Although the capsules produced by this bulk procedure are polydisperse in size and less spherical than those produced using the millifluidic approach, the procedure results in large volumes of dense capsule suspensions. 

\begin{figure}
\includegraphics[width=\columnwidth]{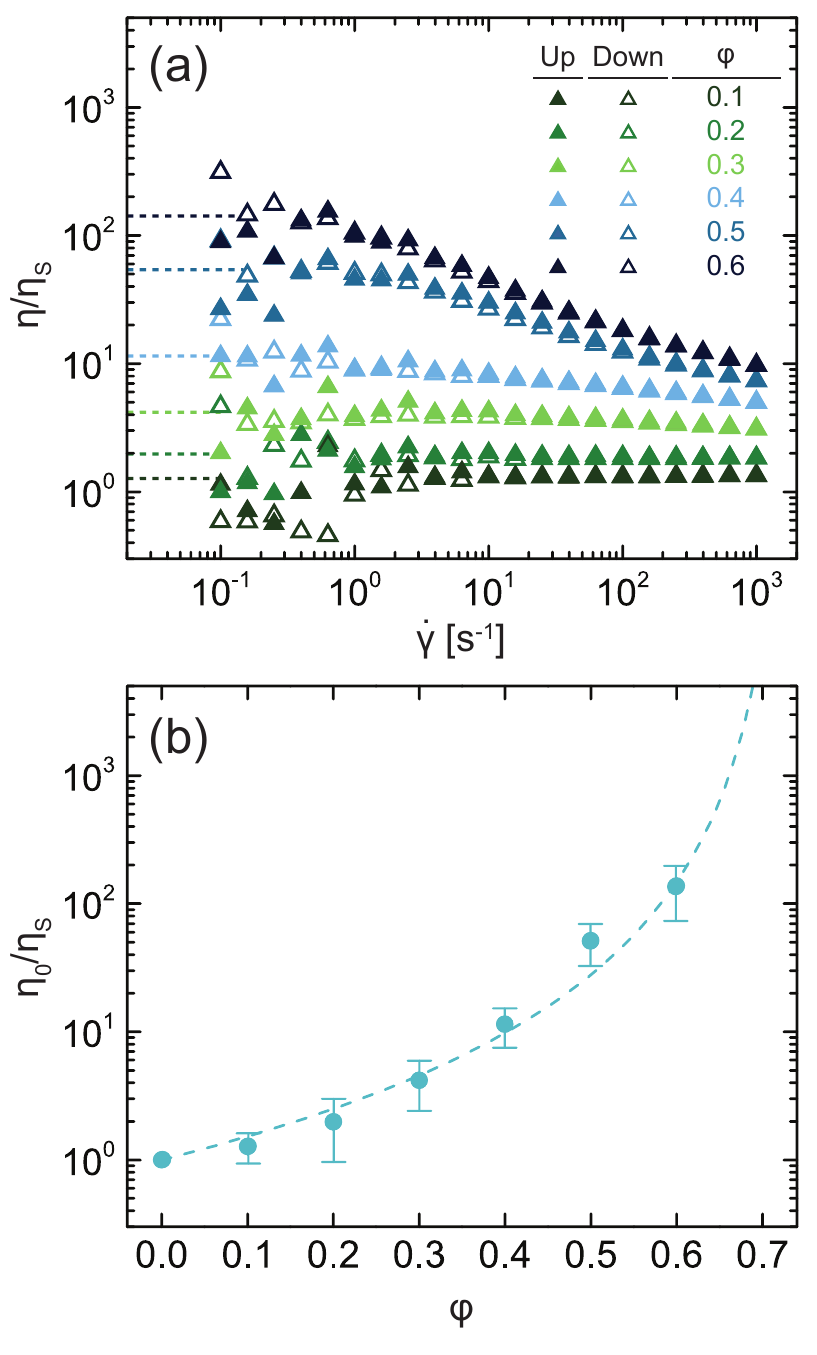}
\caption{(a) Steady shear viscosity $\eta$ normalized to solvent viscosity $\eta_\mathrm{s}$ as a function of shear rate $\dot{\gamma}$ for CNF/OA microcapsule suspensions at various volume fractions $\phi$. Dashed line indicates zero-shear viscosity $\eta_0$. (b) Zero-shear viscosity normalized to the solvent viscosity $\eta_0$/$\eta_S$ of microcapsule suspensions as a function of $\phi$. Dashed curve indicates fit to a modified Krieger-Dougherty equation.}
\label{fig:figure2}
\end{figure}

\subsection{Rheology of microcapsule suspensions}
We measure the viscosity $\eta$ of suspensions with increasing capsule volume fraction $\phi$ in a density-matched solvent mixture of OA, toluene, and chloroform to prevent sedimentation of the capsules (Fig.\ \ref{fig:figure2}). We normalize $\eta$ to that of the solvent $\eta_\mathrm{s}$ to remove inertial effects from the rheometer geometry. At low $\phi$, the suspension viscosity is Newtonian and close to that of the solvent (i.e. $\eta/\eta_\mathrm{s} = 1$). As $\phi$ increases, $\eta/\eta_s$ increases and begins to strongly shear thin with increasing shear rate.  We approximate the zero-shear viscosity $\eta_0$, which characterizes the quiescent suspension properties, by averaging $\eta$ for $\dot{\gamma} < 1$ rad s$^{-1}$. The zero-shear viscosity of these neat microcapsule suspensions can be well described using a modified Krieger-Dougherty equation\cite{Bergstrom1996} $\eta_0/\eta_\mathrm{s} = (1-\phi/\phi_\mathrm{m})^{-[\eta]\phi_\mathrm{m}}$, where $\phi_\mathrm{m} = 0.73 \pm 0.22$ represents the maximum packing fraction and $[\eta] = 4.1 \pm 1.6$ is the intrinsic viscosity of the CNF capsules. These values deviate from predictions for hard spheres and may reflect contributions from particle anisotropy,\cite{Simha1940,Palanisamy2019} polydispersity,\cite{Shewan2014} softness,\cite{Vlassopoulos2014} or weak attractions.\cite{Woutersen1991} This departure from the behavior of hard-sphere suspensions indicates that we may be able to tune the rheology of these suspensions by modifying the capsule properties.

\begin{figure}
\includegraphics[width=\columnwidth]{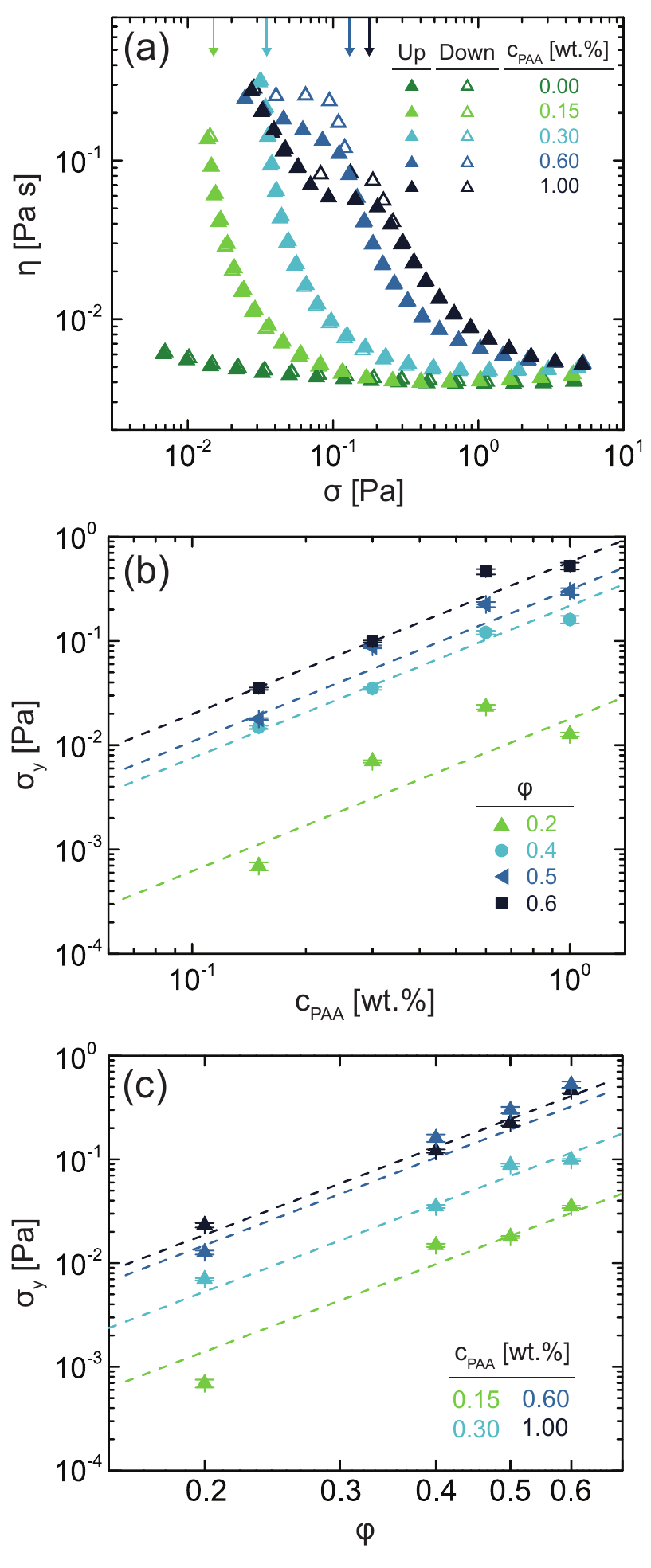}
\caption{(a) Viscosity $\eta$ as a function of shear stress $\sigma$ for microcapsule suspensions at $\phi = 0.4$ and varying PAA concentrations $c_\mathrm{PAA}$. Arrows indicate yield stress values. Yield stress $\sigma_\mathrm{y}$ as a function of (b) PAA concentration $c_\mathrm{PAA}$ and (c) volume fraction $\phi$. Dotted lines represent power-law scalings $\sigma_\mathrm{y} \sim c_\mathrm{PAA}^{1.5}$ and $\sigma_\mathrm{y} \sim \phi^{2.8}$.}
\label{fig:figure3}
\end{figure}

The properties of the capsules can be tuned by incorporating macromolecular additives into the shell. One additive, polyacrylic acid (PAA), has been shown to increase the microcapsule modulus by over an order of magnitude. \cite{Kaufman2017} We incorporate PAA into the microcapsule shells by dissolving it into the CNF suspension before complexation with OA. The rheology of the PAA-modifed capsules exhibits demonstrably different behavior than that of neat capsules. Specifically, $\eta$ diverges at low stresses rather than plateauing, indicating that the material has a characteristic yield stress $\sigma_\mathrm{y}$. The appearance of a yield stress indicates that the PAA-modified microcapsules structure in suspension to form an elastic network under quiescent conditions. Whereas suspensions of non-interacting hard spheres remain predominantly viscous fluids below the glass and jamming transitions, suspensions of attractive colloids develop elasticity at much lower volume fractions as the attractive interactions drive the formation of particle clusters.\cite{Lu2006, Lu2008} Under shear, stress builds within the cluster until it overcomes the attractive interactions, pulls individual colloids apart, and yields the sample.\cite{Sprakel2011} The flow curves for the suspensions of PAA-modified microcapsules (Fig.\ \ref{fig:figure3}(a)) are consistent with this physical picture. At low $\sigma$, $\eta$ diverges because clusters of PAA-modified microcapsules behave as elastic solids. For $\sigma > \sigma_\mathrm{y}$, the suspension viscosity decreases rapidly as the clusters break apart. In the limit of high stress, $\eta$ for the PAA-modified suspensions approaches $\eta$ for suspensions of neat microcapsules because the shear forces completely dominate the interactions between capsules. It is also important to note that the rheology data in the up and down sweeps is highly reproducible indicating that the capsule shells do not break under shear.

\begin{figure}
\includegraphics[width=\columnwidth]{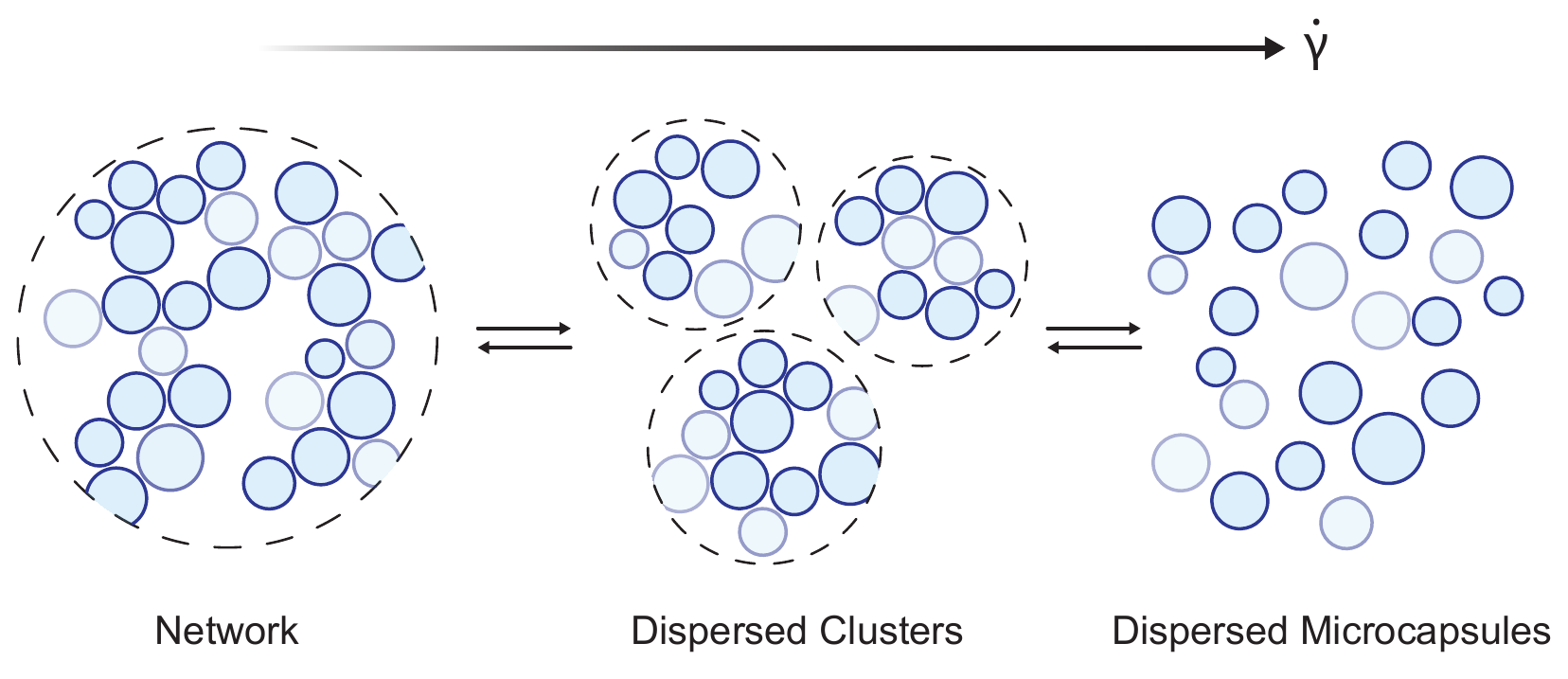}
\caption{Schematic showing suspension microstructure at high $c_\mathrm{PAA}$ as the shear rate $\dot{\gamma}$ (or shear stress $\sigma$) increases.}
\label{fig:figure4}
\end{figure}

At high $c_\mathrm{PAA}$, the microcapsule suspensions exhibit a plateau in the viscosity of the suspension at intermediate stresses and a hysteresis loop between the flow curves with increasing and decreasing shear rates (Fig.\ \ref{fig:figure3}(a)). Hysteresis is commonly observed in structured complex fluids when there is a competition between structural reorganization and the imposed shear rates.\cite{Divoux2013, Radhakrishnan2017} For the suspensions of PAA-modified microcapules, the hysteresis suggests the microcapsules form different structures at intermediate stresses depending on shear history. At low and high $\sigma$, however, the flow curves are independent of shear history as interparticle attractions and shear forces dominate in each regime, respectively. The hysteresis loop occurs at the viscosity plateau for these high $c_\mathrm{PAA}$ suspensions. In attractive suspensions, similar plateaus have been observed for dilute gels\cite{Boromand2017} and for concentrated glasses\cite{Koumakis2011} in which the lower yield stress corresponds to breaking bonds \emph{between} clusters and the stress controlling the end of the viscosity plateau corresponds to breaking bonds \emph{within} clusters. Thus, as the shear rate (or shear stress) increases, the system transitions from a network of percolating clusters, into a suspension of dispersed clusters, and finally into a suspension of individual microcapsules, as shown shown schematically in Fig.\ \ref{fig:figure4}. For this paper, we are interested in how PAA modifies the interparticle interactions and thus focus primarily on the scaling of the higher stress that controls the breaking of interparticle bonds. 

The yield stress for suspensions of attractive colloids is expected to scale as $\sigma_\mathrm{y} \sim U\phi^\delta$, where $U$ represents the magnitude of the interparticle interaction potential and $\delta$ ranges from 1.5 to 6 depending on the system.\cite{Studart2011} To gain insight into how PAA induces attractions between capsules, we extract the yield stress by fitting the flow curves to the Herschel-Bulkley expression $\sigma = \sigma_\mathrm{y} + K\dot{\gamma}^n$, where $K$ is a viscosity prefactor and $n$ is an exponent characterizing the degree of shear thinning. For samples with a viscosity plateau, we extract the stress at the end of the viscosity plateau to compare the interparticle interactions across each sample. The yield stress scales as a power-law with PAA concentration as $\sigma_\mathrm{y} \sim c_\mathrm{PAA}^\alpha$ where $\alpha = 1.5 \pm 0.2$ and with volume fraction as $\sigma_\mathrm{y} \sim \phi^\delta$, where $\delta = 2.8 \pm 0.2$ (Fig.\ \ref{fig:figure3}(b,c)). The volume fraction scaling is in good agreement with literature values for attractive systems.\cite{Roy2020} Given that $\sigma_\mathrm{y}$ scales with $U$, the dependence of $\sigma_\mathrm{y}$ on $c_\mathrm{PAA}$ indicates that $U$ scales with $c_\mathrm{PAA}$. Additionally, the strength of the interparticle attraction increases with increasing polymer concentration as $U \sim c_\mathrm{PAA}^{1/\alpha}$. Thus, incorporating PAA into the microcapsule shell induces strong attractive interactions between capsules with interaction potentials proportional to the concentration of PAA. 

\begin{figure}
\includegraphics[width=\columnwidth]{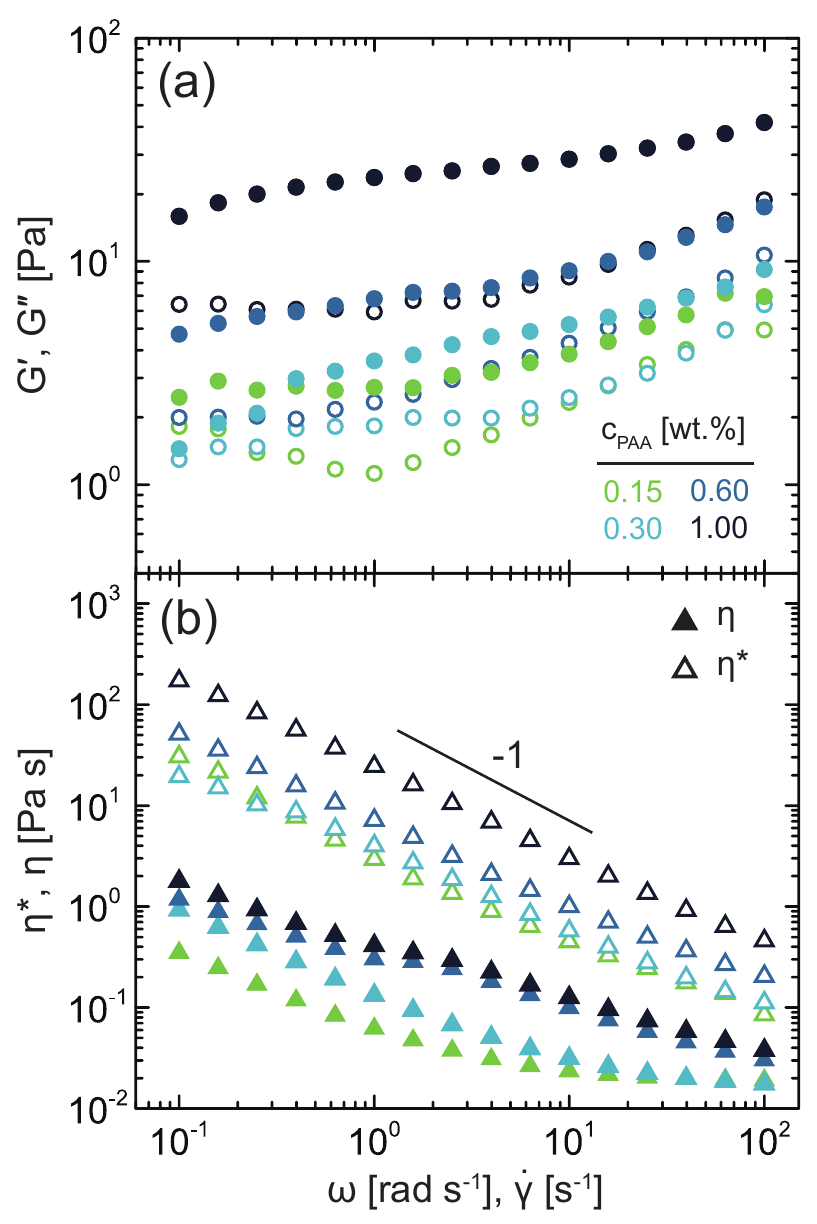}
\caption{(a) Storage ($G'$, closed) and loss ($G''$, open) moduli for suspensions with $\phi = 0.6$ as a function of frequency $\omega$. (b) Complex $\eta^*$ and real $\eta$ viscosity as a function of frequency $\omega$ and shear rate $\dot{\gamma}$, respectively.}
\label{fig:figure5}
\end{figure}

We next use oscillatory rheology to probe the \emph{quiescent} mechanical properties of the network. The suspensions of PAA-modified capsules behave like viscoelastic solids with $G' > G''$ and exhibit only a weak dependence on frequency (Fig.\ \ref{fig:figure5}). As $c_\mathrm{PAA}$ increases, the moduli of the suspensions increase, confirming that the strength of interactions between microcapsule particles increases with increasing $c_\mathrm{PAA}$. We also compare these quiescent properties to the flow properties of the suspensions through the complex $\eta^*$ and real $\eta$ viscosities. In oscillatory and steady shear, the suspension shear thins so that the viscosity decreases with $\omega$ and $\dot{\gamma}$. There is a significant discrepancy, however, between the magnitude of $\eta^*$ and $\eta$ (Fig.\ \ref{fig:figure5}(b)). This discrepancy indicates that steady and oscillatory shear probe different states of the suspension. The steady shear imposed on the suspension of attractive microcapsules disrupts their quiescent structure and thus decreases their contribution to suspension viscosity. By contrast, under oscillatory shear in the linear viscoelastic regime, the suspension maintains its structure so that the rheology reflects the full contribution of the clusters formed by the attractive microcapsules. The comparison between steady and oscillatory shear is fully consistent with the proposed structural changes shown in Fig.\ \ref{fig:figure4}. 

\subsection{Nature of microcapsule interactions}

\begin{figure}
\includegraphics[width=\columnwidth]{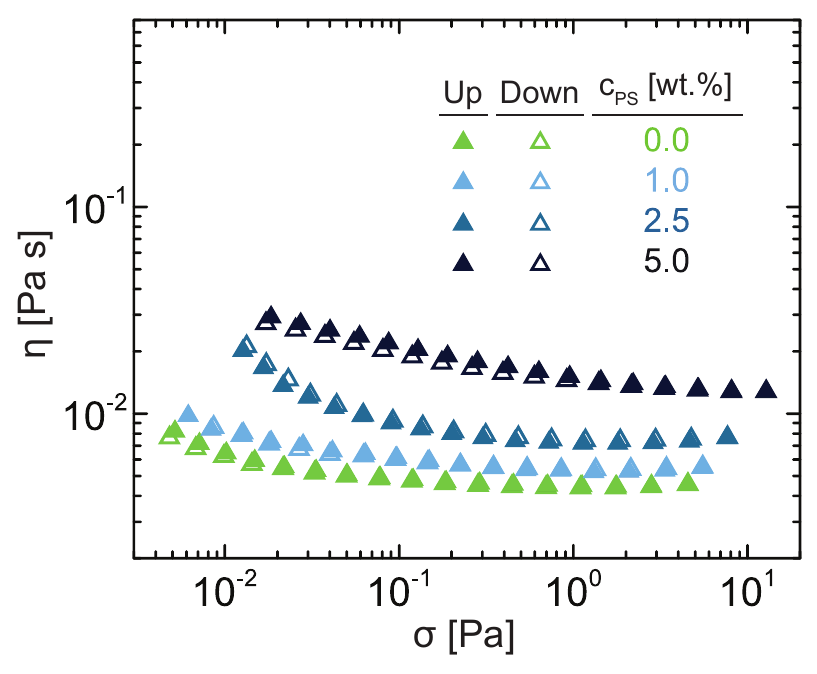}
\caption{Viscosity $\eta$ as a function of shear stress $\sigma$ for neat microcapsules suspended at $\phi = 0.4$ in solutions with varying concentration of polystyrene $c_\mathrm{PS}$.}
\label{fig:figure6}
\end{figure}

The rheological curves measured under steady and oscillatory deformation indicate that the PAA additives induce strong attractions between microcapsules in suspension, but the origin of these attractions remains unclear. In suspensions of hard spheres, attractive interactions are most commonly induced through depletion interactions, in which polymers dissolved in the continuous phase are excluded from regions near the particle surfaces, driving the particles together to minimize the excluded volume.\cite{SaezCabezas2018, Bergenholtz2003, Lu2008} We attempt to induce depletion interactions in suspensions of neat capsules by dissolving polystyrene into the continuous phase and compare the rheological profiles to those for the PAA-modified microcapsule suspensions (Fig.\ \ref{fig:figure6}). The rheology is drastically different between the two systems. In the depletant system, there are minor increases in the suspension viscosity at low stresses (i.e. low shear rates) but no clear divergence indicative of a yield stress. The lack of a clear yield stress indicates that the depletant interactions are much weaker than the PAA-induced attractions. Second, although the depletant system exhibits shear thinning behavior, $\eta$ plateaus at higher values for higher $c_\mathrm{PS}$, in stark contrast to the constant plateau observed in the PAA-system. The different plateaus in the depletant system arise because the dissolved polymer increases the viscosity of the continuous phase surrounding the microcapsules. This effect is not observed in the PAA-modified suspensions because the PAA is expected to remain encapsulated within the CNF shell. These differences in the rheology of the two classes of suspensions indicates that the attractive interactions in the PAA-modified suspension are not depletant-like.

We hypothesize instead that the PAA induces interactions between the CNF microcapsules by changing the surface chemistry of the shell resulting in capsules sticking to each other. It is well-known that PAA dissolved in the continuous phase bridges particles to induce attractive interaction and flocculation in aqueous suspensions.\cite{Yu1996, Biggs1995} For the PAA-modified microcapsule suspensions, however, the continuous phase is organic rather than aqueous, making it highly unlikely that a significant fraction of PAA is dispersed in the continuous phase. Instead, we believe the PAA is incorporated into the porous CNF shell to modify the shell chemistry. Thus, by encapsulating PAA in the internal phase of the capsule, we modify the shell properties and induce interactions between capsules in suspension. These sticky interactions require direct capsule contact to operate. We expect that the kinetics of these sticky attractions compete with the breaking apart of clusters under shear to generate the hysteresis and plateau viscosity observed under flow.

\section{Conclusions}
We produce microcapsules approximately 25-200 $\mu$m in diameter through a bulk emulsification-based method in which cellulose nanofibrils complex with oleylamine at the oil-water interface. The resulting microcapsules are porous to allow exchange with their environment and are mechanically robust enough to survive processing and solvent transfer. We control the suspension rheology by incorporating polyacrylic acid into the microcapsule shell, which induces strong attractions between suspended capsules and leads to the emergence of a yield stress and hysteresis in the flow curves. The yield stress $\sigma_\mathrm{y}$ increases as a power-law with suspension volume fraction $\phi$ and PAA concentration $c_\mathrm{PAA}$. Additionally, the attractions between capsules results in a strong elastic network whose modulus increases with increasing $c_\mathrm{PAA}$. These rheological signatures suggest that the PAA-modified capsules form clusters in suspension that break apart under shear stress.

Importantly, we demonstrate that the rheological behavior of microcapsule suspensions can be modified through additives to the \emph{internal} phase of the capsules. These internal additives do not affect the properties of the continuous phase but instead generate attractive interactions between capsules to control the overall macroscopic rheology. We expect this finding to be particularly useful when designing injectable suspensions. Injectable suspensions require strong shear thinning to pass through a syringe but then must recover elasticity to remain in place.\cite{Daly2020} Changes to the shell chemistry are controlled by the composition of the internal phase. By modifying the rheological properties in this fashion, the properties of these suspensions may be more stable against changes in the local environment. 

\section{Acknowledgments}
We thank Uri Gabinet and Dr. Gilad Kaufman for helpful discussions. This research was supported by the Center for Engineering MechanoBiology, an NSF Science and Technology Center, under grant agreement CMMI: 15-48571.

\bibliography{biblio}

\begin{thebibliography}{53}%
\makeatletter
\providecommand \@ifxundefined [1]{%
 \@ifx{#1\undefined}
}%
\providecommand \@ifnum [1]{%
 \ifnum #1\expandafter \@firstoftwo
 \else \expandafter \@secondoftwo
 \fi
}%
\providecommand \@ifx [1]{%
 \ifx #1\expandafter \@firstoftwo
 \else \expandafter \@secondoftwo
 \fi
}%
\providecommand \natexlab [1]{#1}%
\providecommand \enquote  [1]{``#1''}%
\providecommand \bibnamefont  [1]{#1}%
\providecommand \bibfnamefont [1]{#1}%
\providecommand \citenamefont [1]{#1}%
\providecommand \href@noop [0]{\@secondoftwo}%
\providecommand \href [0]{\begingroup \@sanitize@url \@href}%
\providecommand \@href[1]{\@@startlink{#1}\@@href}%
\providecommand \@@href[1]{\endgroup#1\@@endlink}%
\providecommand \@sanitize@url [0]{\catcode `\\12\catcode `\$12\catcode
  `\&12\catcode `\#12\catcode `\^12\catcode `\_12\catcode `\%12\relax}%
\providecommand \@@startlink[1]{}%
\providecommand \@@endlink[0]{}%
\providecommand \url  [0]{\begingroup\@sanitize@url \@url }%
\providecommand \@url [1]{\endgroup\@href {#1}{\urlprefix }}%
\providecommand \urlprefix  [0]{URL }%
\providecommand \Eprint [0]{\href }%
\providecommand \doibase [0]{https://doi.org/}%
\providecommand \selectlanguage [0]{\@gobble}%
\providecommand \bibinfo  [0]{\@secondoftwo}%
\providecommand \bibfield  [0]{\@secondoftwo}%
\providecommand \translation [1]{[#1]}%
\providecommand \BibitemOpen [0]{}%
\providecommand \bibitemStop [0]{}%
\providecommand \bibitemNoStop [0]{.\EOS\space}%
\providecommand \EOS [0]{\spacefactor3000\relax}%
\providecommand \BibitemShut  [1]{\csname bibitem#1\endcsname}%
\let\auto@bib@innerbib\@empty
\bibitem [{\citenamefont {Yow}\ and\ \citenamefont {Routh}(2006)}]{Yow2006}%
  \BibitemOpen
  \bibfield  {author} {\bibinfo {author} {\bibfnamefont {H.~N.}\ \bibnamefont
  {Yow}}\ and\ \bibinfo {author} {\bibfnamefont {A.~F.}\ \bibnamefont
  {Routh}},\ }\bibfield  {title} {\bibinfo {title} {{Formation of liquid
  core-polymer shell microcapsules}},\ }\href
  {https://doi.org/10.1039/b606965g} {\bibfield  {journal} {\bibinfo  {journal}
  {Soft Matter}\ }\textbf {\bibinfo {volume} {2}},\ \bibinfo {pages} {940}
  (\bibinfo {year} {2006})}\BibitemShut {NoStop}%
\bibitem [{\citenamefont {Lensen}\ \emph {et~al.}(2008)\citenamefont {Lensen},
  \citenamefont {Vriezema},\ and\ \citenamefont {van Hest}}]{Lensen2008}%
  \BibitemOpen
  \bibfield  {author} {\bibinfo {author} {\bibfnamefont {D.}~\bibnamefont
  {Lensen}}, \bibinfo {author} {\bibfnamefont {D.~M.}\ \bibnamefont
  {Vriezema}},\ and\ \bibinfo {author} {\bibfnamefont {J.~C.}\ \bibnamefont
  {van Hest}},\ }\bibfield  {title} {\bibinfo {title} {{Polymeric microcapsules
  for synthetic applications}},\ }\href
  {https://doi.org/10.1002/mabi.200800112} {\bibfield  {journal} {\bibinfo
  {journal} {Macromol. Biosci.}\ }\textbf {\bibinfo {volume} {8}},\ \bibinfo
  {pages} {991} (\bibinfo {year} {2008})}\BibitemShut {NoStop}%
\bibitem [{\citenamefont {Amstad}(2017)}]{Amstad2017}%
  \BibitemOpen
  \bibfield  {author} {\bibinfo {author} {\bibfnamefont {E.}~\bibnamefont
  {Amstad}},\ }\bibfield  {title} {\bibinfo {title} {{Capsules: Their past and
  opportunities for their future}},\ }\href
  {https://doi.org/10.1021/acsmacrolett.7b00472} {\bibfield  {journal}
  {\bibinfo  {journal} {ACS Macro Lett.}\ }\textbf {\bibinfo {volume} {6}},\
  \bibinfo {pages} {841} (\bibinfo {year} {2017})}\BibitemShut {NoStop}%
\bibitem [{\citenamefont {Zhang}\ \emph {et~al.}(2012)\citenamefont {Zhang},
  \citenamefont {Coulston}, \citenamefont {Jones}, \citenamefont {Geng},
  \citenamefont {Scherman},\ and\ \citenamefont {Abell}}]{Zhang2012}%
  \BibitemOpen
  \bibfield  {author} {\bibinfo {author} {\bibfnamefont {J.}~\bibnamefont
  {Zhang}}, \bibinfo {author} {\bibfnamefont {R.~J.}\ \bibnamefont {Coulston}},
  \bibinfo {author} {\bibfnamefont {S.~T.}\ \bibnamefont {Jones}}, \bibinfo
  {author} {\bibfnamefont {J.}~\bibnamefont {Geng}}, \bibinfo {author}
  {\bibfnamefont {O.~A.}\ \bibnamefont {Scherman}},\ and\ \bibinfo {author}
  {\bibfnamefont {C.}~\bibnamefont {Abell}},\ }\bibfield  {title} {\bibinfo
  {title} {{One-Step Fabrication of Supramolecular Microcapsules from
  Microfluidic Droplets}},\ }\href {https://doi.org/10.1126/science.1215416}
  {\bibfield  {journal} {\bibinfo  {journal} {Science.}\ }\textbf {\bibinfo
  {volume} {335}},\ \bibinfo {pages} {690} (\bibinfo {year}
  {2012})}\BibitemShut {NoStop}%
\bibitem [{\citenamefont {Kaufman}\ \emph {et~al.}(2014)\citenamefont
  {Kaufman}, \citenamefont {Boltyanskiy}, \citenamefont {Nejati}, \citenamefont
  {Thiam}, \citenamefont {Loewenberg}, \citenamefont {Dufresne},\ and\
  \citenamefont {Osuji}}]{Kaufman2014}%
  \BibitemOpen
  \bibfield  {author} {\bibinfo {author} {\bibfnamefont {G.}~\bibnamefont
  {Kaufman}}, \bibinfo {author} {\bibfnamefont {R.}~\bibnamefont
  {Boltyanskiy}}, \bibinfo {author} {\bibfnamefont {S.}~\bibnamefont {Nejati}},
  \bibinfo {author} {\bibfnamefont {A.~R.}\ \bibnamefont {Thiam}}, \bibinfo
  {author} {\bibfnamefont {M.}~\bibnamefont {Loewenberg}}, \bibinfo {author}
  {\bibfnamefont {E.~R.}\ \bibnamefont {Dufresne}},\ and\ \bibinfo {author}
  {\bibfnamefont {C.~O.}\ \bibnamefont {Osuji}},\ }\bibfield  {title} {\bibinfo
  {title} {{Single-step microfluidic fabrication of soft monodisperse
  polyelectrolyte microcapsules by interfacial complexation}},\ }\href
  {https://doi.org/10.1039/c4lc00482e} {\bibfield  {journal} {\bibinfo
  {journal} {Lab Chip}\ }\textbf {\bibinfo {volume} {14}},\ \bibinfo {pages}
  {3494} (\bibinfo {year} {2014})}\BibitemShut {NoStop}%
\bibitem [{\citenamefont {Kaufman}\ \emph {et~al.}(2017)\citenamefont
  {Kaufman}, \citenamefont {Mukhopadhyay}, \citenamefont {Rokhlenko},
  \citenamefont {Nejati}, \citenamefont {Boltyanskiy}, \citenamefont {Choo},
  \citenamefont {Loewenberg},\ and\ \citenamefont {Osuji}}]{Kaufman2017}%
  \BibitemOpen
  \bibfield  {author} {\bibinfo {author} {\bibfnamefont {G.}~\bibnamefont
  {Kaufman}}, \bibinfo {author} {\bibfnamefont {S.}~\bibnamefont
  {Mukhopadhyay}}, \bibinfo {author} {\bibfnamefont {Y.}~\bibnamefont
  {Rokhlenko}}, \bibinfo {author} {\bibfnamefont {S.}~\bibnamefont {Nejati}},
  \bibinfo {author} {\bibfnamefont {R.}~\bibnamefont {Boltyanskiy}}, \bibinfo
  {author} {\bibfnamefont {Y.}~\bibnamefont {Choo}}, \bibinfo {author}
  {\bibfnamefont {M.}~\bibnamefont {Loewenberg}},\ and\ \bibinfo {author}
  {\bibfnamefont {C.~O.}\ \bibnamefont {Osuji}},\ }\bibfield  {title} {\bibinfo
  {title} {{Highly stiff yet elastic microcapsules incorporating cellulose
  nanofibrils}},\ }\href {https://doi.org/10.1039/C7SM00092H} {\bibfield
  {journal} {\bibinfo  {journal} {Soft Matter}\ }\textbf {\bibinfo {volume}
  {13}},\ \bibinfo {pages} {2733} (\bibinfo {year} {2017})}\BibitemShut
  {NoStop}%
\bibitem [{\citenamefont {{Dupr{\'{e}} De Baubigny}}\ \emph
  {et~al.}(2017)\citenamefont {{Dupr{\'{e}} De Baubigny}}, \citenamefont
  {Tr{\'{e}}gou{\"{e}}t}, \citenamefont {Salez}, \citenamefont {Pantoustier},
  \citenamefont {Perrin}, \citenamefont {Reyssat},\ and\ \citenamefont
  {Monteux}}]{DupreDeBaubigny2017}%
  \BibitemOpen
  \bibfield  {author} {\bibinfo {author} {\bibfnamefont {J.}~\bibnamefont
  {{Dupr{\'{e}} De Baubigny}}}, \bibinfo {author} {\bibfnamefont
  {C.}~\bibnamefont {Tr{\'{e}}gou{\"{e}}t}}, \bibinfo {author} {\bibfnamefont
  {T.}~\bibnamefont {Salez}}, \bibinfo {author} {\bibfnamefont
  {N.}~\bibnamefont {Pantoustier}}, \bibinfo {author} {\bibfnamefont
  {P.}~\bibnamefont {Perrin}}, \bibinfo {author} {\bibfnamefont
  {M.}~\bibnamefont {Reyssat}},\ and\ \bibinfo {author} {\bibfnamefont
  {C.}~\bibnamefont {Monteux}},\ }\bibfield  {title} {\bibinfo {title}
  {{One-Step Fabrication of pH-Responsive Membranes and Microcapsules through
  Interfacial H-Bond Polymer Complexation}},\ }\href
  {https://doi.org/10.1038/s41598-017-01374-3} {\bibfield  {journal} {\bibinfo
  {journal} {Sci. Rep.}\ }\textbf {\bibinfo {volume} {7}},\ \bibinfo {pages}
  {1265} (\bibinfo {year} {2017})}\BibitemShut {NoStop}%
\bibitem [{\citenamefont {Song}\ \emph {et~al.}(2019)\citenamefont {Song},
  \citenamefont {Babayekhorasani},\ and\ \citenamefont {Spicer}}]{Song2019}%
  \BibitemOpen
  \bibfield  {author} {\bibinfo {author} {\bibfnamefont {J.}~\bibnamefont
  {Song}}, \bibinfo {author} {\bibfnamefont {F.}~\bibnamefont
  {Babayekhorasani}},\ and\ \bibinfo {author} {\bibfnamefont {P.~T.}\
  \bibnamefont {Spicer}},\ }\bibfield  {title} {\bibinfo {title} {{Soft
  Bacterial Cellulose Microcapsules with Adaptable Shapes}},\ }\href
  {https://doi.org/10.1021/acs.biomac.9b01143} {\bibfield  {journal} {\bibinfo
  {journal} {Biomacromolecules}\ }\textbf {\bibinfo {volume} {20}},\ \bibinfo
  {pages} {4437} (\bibinfo {year} {2019})}\BibitemShut {NoStop}%
\bibitem [{\citenamefont {Liu}\ \emph {et~al.}(2020)\citenamefont {Liu},
  \citenamefont {Li}, \citenamefont {Xia}, \citenamefont {Zhu}, \citenamefont
  {Chen},\ and\ \citenamefont {Chen}}]{Liu2020}%
  \BibitemOpen
  \bibfield  {author} {\bibinfo {author} {\bibfnamefont {Z.}~\bibnamefont
  {Liu}}, \bibinfo {author} {\bibfnamefont {S.}~\bibnamefont {Li}}, \bibinfo
  {author} {\bibfnamefont {X.}~\bibnamefont {Xia}}, \bibinfo {author}
  {\bibfnamefont {Z.}~\bibnamefont {Zhu}}, \bibinfo {author} {\bibfnamefont
  {L.}~\bibnamefont {Chen}},\ and\ \bibinfo {author} {\bibfnamefont
  {Z.}~\bibnamefont {Chen}},\ }\bibfield  {title} {\bibinfo {title} {{Recent
  Advances in Multifunctional Graphitic Nanocapsules for Raman Detection,
  Imaging, and Therapy}},\ }\href {https://doi.org/10.1002/smtd.201900440}
  {\bibfield  {journal} {\bibinfo  {journal} {Small Methods}\ }\textbf
  {\bibinfo {volume} {4}},\ \bibinfo {pages} {1900440} (\bibinfo {year}
  {2020})}\BibitemShut {NoStop}%
\bibitem [{\citenamefont {Dinsmore}\ \emph {et~al.}(2002)\citenamefont
  {Dinsmore}, \citenamefont {Hsu}, \citenamefont {Nikolaides}, \citenamefont
  {Marquez}, \citenamefont {Bausch},\ and\ \citenamefont
  {Weitz}}]{Dinsmore2002}%
  \BibitemOpen
  \bibfield  {author} {\bibinfo {author} {\bibfnamefont {A.~D.}\ \bibnamefont
  {Dinsmore}}, \bibinfo {author} {\bibfnamefont {M.~F.}\ \bibnamefont {Hsu}},
  \bibinfo {author} {\bibfnamefont {M.~G.}\ \bibnamefont {Nikolaides}},
  \bibinfo {author} {\bibfnamefont {M.}~\bibnamefont {Marquez}}, \bibinfo
  {author} {\bibfnamefont {A.~R.}\ \bibnamefont {Bausch}},\ and\ \bibinfo
  {author} {\bibfnamefont {D.~A.}\ \bibnamefont {Weitz}},\ }\bibfield  {title}
  {\bibinfo {title} {{Colloidosomes: Selectively permeable capsules composed of
  colloidal particles}},\ }\href {https://doi.org/10.1126/science.1074868}
  {\bibfield  {journal} {\bibinfo  {journal} {Science}\ }\textbf {\bibinfo
  {volume} {298}},\ \bibinfo {pages} {1006} (\bibinfo {year}
  {2002})}\BibitemShut {NoStop}%
\bibitem [{\citenamefont {Discher}\ \emph {et~al.}(1999)\citenamefont
  {Discher}, \citenamefont {Won}, \citenamefont {Ege}, \citenamefont {Lee},
  \citenamefont {Bates}, \citenamefont {Discher},\ and\ \citenamefont
  {Hammer}}]{Discher1999}%
  \BibitemOpen
  \bibfield  {author} {\bibinfo {author} {\bibfnamefont {B.~M.}\ \bibnamefont
  {Discher}}, \bibinfo {author} {\bibfnamefont {Y.~Y.}\ \bibnamefont {Won}},
  \bibinfo {author} {\bibfnamefont {D.~S.}\ \bibnamefont {Ege}}, \bibinfo
  {author} {\bibfnamefont {J.~C.}\ \bibnamefont {Lee}}, \bibinfo {author}
  {\bibfnamefont {F.~S.}\ \bibnamefont {Bates}}, \bibinfo {author}
  {\bibfnamefont {D.~E.}\ \bibnamefont {Discher}},\ and\ \bibinfo {author}
  {\bibfnamefont {D.~A.}\ \bibnamefont {Hammer}},\ }\bibfield  {title}
  {\bibinfo {title} {{Polymersomes: Tough vesicles made from diblock
  copolymers}},\ }\href {https://doi.org/10.1126/science.284.5417.1143}
  {\bibfield  {journal} {\bibinfo  {journal} {Science}\ }\textbf {\bibinfo
  {volume} {284}},\ \bibinfo {pages} {1143} (\bibinfo {year}
  {1999})}\BibitemShut {NoStop}%
\bibitem [{\citenamefont {B{\'{e}}dard}\ \emph {et~al.}(2010)\citenamefont
  {B{\'{e}}dard}, \citenamefont {{De Geest}}, \citenamefont {Skirtach},
  \citenamefont {M{\"{o}}hwald},\ and\ \citenamefont
  {Sukhorukov}}]{Bedard2010}%
  \BibitemOpen
  \bibfield  {author} {\bibinfo {author} {\bibfnamefont {M.~F.}\ \bibnamefont
  {B{\'{e}}dard}}, \bibinfo {author} {\bibfnamefont {B.~G.}\ \bibnamefont {{De
  Geest}}}, \bibinfo {author} {\bibfnamefont {A.~G.}\ \bibnamefont {Skirtach}},
  \bibinfo {author} {\bibfnamefont {H.}~\bibnamefont {M{\"{o}}hwald}},\ and\
  \bibinfo {author} {\bibfnamefont {G.~B.}\ \bibnamefont {Sukhorukov}},\
  }\bibfield  {title} {\bibinfo {title} {{Polymeric microcapsules with light
  responsive properties for encapsulation and release}},\ }\href
  {https://doi.org/10.1016/j.cis.2009.07.007} {\bibfield  {journal} {\bibinfo
  {journal} {Adv. Colloid Interface Sci.}\ }\textbf {\bibinfo {volume} {158}},\
  \bibinfo {pages} {2} (\bibinfo {year} {2010})}\BibitemShut {NoStop}%
\bibitem [{\citenamefont {Blaiszik}\ \emph {et~al.}(2010)\citenamefont
  {Blaiszik}, \citenamefont {Kramer}, \citenamefont {Olugebefola},
  \citenamefont {Moore}, \citenamefont {Sottos},\ and\ \citenamefont
  {White}}]{Blaiszik2010}%
  \BibitemOpen
  \bibfield  {author} {\bibinfo {author} {\bibfnamefont {B.}~\bibnamefont
  {Blaiszik}}, \bibinfo {author} {\bibfnamefont {S.}~\bibnamefont {Kramer}},
  \bibinfo {author} {\bibfnamefont {S.}~\bibnamefont {Olugebefola}}, \bibinfo
  {author} {\bibfnamefont {J.}~\bibnamefont {Moore}}, \bibinfo {author}
  {\bibfnamefont {N.}~\bibnamefont {Sottos}},\ and\ \bibinfo {author}
  {\bibfnamefont {S.}~\bibnamefont {White}},\ }\bibfield  {title} {\bibinfo
  {title} {{Self-Healing Polymers and Composites}},\ }\href
  {https://doi.org/10.1146/annurev-matsci-070909-104532} {\bibfield  {journal}
  {\bibinfo  {journal} {Annu. Rev. Mater. Res.}\ }\textbf {\bibinfo {volume}
  {40}},\ \bibinfo {pages} {179} (\bibinfo {year} {2010})}\BibitemShut
  {NoStop}%
\bibitem [{\citenamefont {Morimoto}\ \emph {et~al.}(2009)\citenamefont
  {Morimoto}, \citenamefont {Tan}, \citenamefont {Tsuda},\ and\ \citenamefont
  {Takeuchi}}]{Morimoto2009}%
  \BibitemOpen
  \bibfield  {author} {\bibinfo {author} {\bibfnamefont {Y.}~\bibnamefont
  {Morimoto}}, \bibinfo {author} {\bibfnamefont {W.~H.}\ \bibnamefont {Tan}},
  \bibinfo {author} {\bibfnamefont {Y.}~\bibnamefont {Tsuda}},\ and\ \bibinfo
  {author} {\bibfnamefont {S.}~\bibnamefont {Takeuchi}},\ }\bibfield  {title}
  {\bibinfo {title} {{Monodisperse semi-permeable microcapsules for continuous
  observation of cells}},\ }\href {https://doi.org/10.1039/b900035f} {\bibfield
   {journal} {\bibinfo  {journal} {Lab Chip}\ }\textbf {\bibinfo {volume}
  {9}},\ \bibinfo {pages} {2217} (\bibinfo {year} {2009})}\BibitemShut
  {NoStop}%
\bibitem [{\citenamefont {Niepa}\ \emph {et~al.}(2016)\citenamefont {Niepa},
  \citenamefont {Hou}, \citenamefont {Jiang}, \citenamefont {Goulian},
  \citenamefont {Koo}, \citenamefont {Stebe},\ and\ \citenamefont
  {Lee}}]{Niepa2016}%
  \BibitemOpen
  \bibfield  {author} {\bibinfo {author} {\bibfnamefont {T.~H.~R.}\
  \bibnamefont {Niepa}}, \bibinfo {author} {\bibfnamefont {L.}~\bibnamefont
  {Hou}}, \bibinfo {author} {\bibfnamefont {H.}~\bibnamefont {Jiang}}, \bibinfo
  {author} {\bibfnamefont {M.}~\bibnamefont {Goulian}}, \bibinfo {author}
  {\bibfnamefont {H.}~\bibnamefont {Koo}}, \bibinfo {author} {\bibfnamefont
  {K.~J.}\ \bibnamefont {Stebe}},\ and\ \bibinfo {author} {\bibfnamefont
  {D.}~\bibnamefont {Lee}},\ }\bibfield  {title} {\bibinfo {title} {{Microbial
  Nanoculture as an Artificial Microniche}},\ }\href
  {https://doi.org/10.1038/srep30578} {\bibfield  {journal} {\bibinfo
  {journal} {Sci. Rep.}\ }\textbf {\bibinfo {volume} {6}},\ \bibinfo {pages}
  {30578} (\bibinfo {year} {2016})}\BibitemShut {NoStop}%
\bibitem [{\citenamefont {Manimaran}\ \emph {et~al.}(2020)\citenamefont
  {Manimaran}, \citenamefont {Usman}, \citenamefont {Kamga}, \citenamefont
  {Davidson}, \citenamefont {Beckman},\ and\ \citenamefont
  {Niepa}}]{Manimaran2020}%
  \BibitemOpen
  \bibfield  {author} {\bibinfo {author} {\bibfnamefont {N.~H.}\ \bibnamefont
  {Manimaran}}, \bibinfo {author} {\bibfnamefont {H.}~\bibnamefont {Usman}},
  \bibinfo {author} {\bibfnamefont {K.~L.}\ \bibnamefont {Kamga}}, \bibinfo
  {author} {\bibfnamefont {S.-l.}\ \bibnamefont {Davidson}}, \bibinfo {author}
  {\bibfnamefont {E.}~\bibnamefont {Beckman}},\ and\ \bibinfo {author}
  {\bibfnamefont {T.~H.~R.}\ \bibnamefont {Niepa}},\ }\bibfield  {title}
  {\bibinfo {title} {{Developing a Functional Poly(dimethylsiloxane)-Based
  Microbial Nanoculture System Using Dimethylallylamine}},\ }\href
  {https://doi.org/10.1021/acsami.0c11875} {\bibfield  {journal} {\bibinfo
  {journal} {ACS Appl. Mater. Interfaces}\ }\textbf {\bibinfo {volume} {12}},\
  \bibinfo {pages} {50581} (\bibinfo {year} {2020})}\BibitemShut {NoStop}%
\bibitem [{\citenamefont {Mewis}\ and\ \citenamefont
  {Wagner}(2012)}]{Mewis2012}%
  \BibitemOpen
  \bibfield  {author} {\bibinfo {author} {\bibfnamefont {J.}~\bibnamefont
  {Mewis}}\ and\ \bibinfo {author} {\bibfnamefont {N.~J.}\ \bibnamefont
  {Wagner}},\ }\href@noop {} {\emph {\bibinfo {title} {{Colloidal suspension
  rheology}}}}\ (\bibinfo  {publisher} {Cambridge University Press},\ \bibinfo
  {year} {2012})\BibitemShut {NoStop}%
\bibitem [{\citenamefont {Wildemuth}\ and\ \citenamefont
  {Williams}(1984)}]{Wildemuth1984}%
  \BibitemOpen
  \bibfield  {author} {\bibinfo {author} {\bibfnamefont {C.~R.}\ \bibnamefont
  {Wildemuth}}\ and\ \bibinfo {author} {\bibfnamefont {M.~C.}\ \bibnamefont
  {Williams}},\ }\bibfield  {title} {\bibinfo {title} {{Viscosity of
  suspensions modeled with a shear-dependent maximum packing fraction}},\
  }\href {https://doi.org/10.1007/BF01438803} {\bibfield  {journal} {\bibinfo
  {journal} {Rheol. Acta}\ }\textbf {\bibinfo {volume} {23}},\ \bibinfo {pages}
  {627} (\bibinfo {year} {1984})}\BibitemShut {NoStop}%
\bibitem [{\citenamefont {Hunter}\ and\ \citenamefont
  {Weeks}(2012)}]{Hunter2012}%
  \BibitemOpen
  \bibfield  {author} {\bibinfo {author} {\bibfnamefont {G.~L.}\ \bibnamefont
  {Hunter}}\ and\ \bibinfo {author} {\bibfnamefont {E.~R.}\ \bibnamefont
  {Weeks}},\ }\bibfield  {title} {\bibinfo {title} {{The physics of the
  colloidal glass transition}},\ }\href
  {https://doi.org/10.1088/0034-4885/75/6/066501} {\bibfield  {journal}
  {\bibinfo  {journal} {Reports Prog. Phys.}\ }\textbf {\bibinfo {volume}
  {75}},\ \bibinfo {pages} {066501} (\bibinfo {year} {2012})}\BibitemShut
  {NoStop}%
\bibitem [{\citenamefont {Liu}\ \emph {et~al.}(1996)\citenamefont {Liu},
  \citenamefont {Williams},\ and\ \citenamefont {Briscoe}}]{Liu1996}%
  \BibitemOpen
  \bibfield  {author} {\bibinfo {author} {\bibfnamefont {K.~K.}\ \bibnamefont
  {Liu}}, \bibinfo {author} {\bibfnamefont {D.~R.}\ \bibnamefont {Williams}},\
  and\ \bibinfo {author} {\bibfnamefont {B.~J.}\ \bibnamefont {Briscoe}},\
  }\bibfield  {title} {\bibinfo {title} {{Compressive deformation of a single
  microcapsule}},\ }\href {https://doi.org/10.1103/PhysRevE.54.6673} {\bibfield
   {journal} {\bibinfo  {journal} {Phys. Rev. E}\ }\textbf {\bibinfo {volume}
  {54}},\ \bibinfo {pages} {6673} (\bibinfo {year} {1996})}\BibitemShut
  {NoStop}%
\bibitem [{\citenamefont {Vlassopoulos}\ and\ \citenamefont
  {Cloitre}(2014)}]{Vlassopoulos2014}%
  \BibitemOpen
  \bibfield  {author} {\bibinfo {author} {\bibfnamefont {D.}~\bibnamefont
  {Vlassopoulos}}\ and\ \bibinfo {author} {\bibfnamefont {M.}~\bibnamefont
  {Cloitre}},\ }\bibfield  {title} {\bibinfo {title} {{Tunable rheology of
  dense soft deformable colloids}},\ }\href
  {https://doi.org/10.1016/j.cocis.2014.09.007} {\bibfield  {journal} {\bibinfo
   {journal} {Curr. Opin. Colloid Interface Sci.}\ }\textbf {\bibinfo {volume}
  {19}},\ \bibinfo {pages} {561} (\bibinfo {year} {2014})}\BibitemShut
  {NoStop}%
\bibitem [{\citenamefont {Mattsson}\ \emph {et~al.}(2009)\citenamefont
  {Mattsson}, \citenamefont {Wyss}, \citenamefont {Fernandez-Nieves},
  \citenamefont {Miyazaki}, \citenamefont {Hu}, \citenamefont {Reichman},\ and\
  \citenamefont {Weitz}}]{Mattsson2009}%
  \BibitemOpen
  \bibfield  {author} {\bibinfo {author} {\bibfnamefont {J.}~\bibnamefont
  {Mattsson}}, \bibinfo {author} {\bibfnamefont {H.~M.}\ \bibnamefont {Wyss}},
  \bibinfo {author} {\bibfnamefont {A.}~\bibnamefont {Fernandez-Nieves}},
  \bibinfo {author} {\bibfnamefont {K.}~\bibnamefont {Miyazaki}}, \bibinfo
  {author} {\bibfnamefont {Z.}~\bibnamefont {Hu}}, \bibinfo {author}
  {\bibfnamefont {D.~R.}\ \bibnamefont {Reichman}},\ and\ \bibinfo {author}
  {\bibfnamefont {D.~A.}\ \bibnamefont {Weitz}},\ }\bibfield  {title} {\bibinfo
  {title} {{Soft colloids make strong glasses}},\ }\href
  {https://doi.org/10.1038/nature08457} {\bibfield  {journal} {\bibinfo
  {journal} {Nature}\ }\textbf {\bibinfo {volume} {462}},\ \bibinfo {pages}
  {83} (\bibinfo {year} {2009})}\BibitemShut {NoStop}%
\bibitem [{\citenamefont {Pellet}\ and\ \citenamefont
  {Cloitre}(2016)}]{Pellet2016}%
  \BibitemOpen
  \bibfield  {author} {\bibinfo {author} {\bibfnamefont {C.}~\bibnamefont
  {Pellet}}\ and\ \bibinfo {author} {\bibfnamefont {M.}~\bibnamefont
  {Cloitre}},\ }\bibfield  {title} {\bibinfo {title} {{The glass and jamming
  transitions of soft polyelectrolyte microgel suspensions}},\ }\href
  {https://doi.org/10.1039/c5sm03001c} {\bibfield  {journal} {\bibinfo
  {journal} {Soft Matter}\ }\textbf {\bibinfo {volume} {12}},\ \bibinfo {pages}
  {3710} (\bibinfo {year} {2016})}\BibitemShut {NoStop}%
\bibitem [{\citenamefont {Gnan}\ and\ \citenamefont
  {Zaccarelli}(2019)}]{Gnan2019}%
  \BibitemOpen
  \bibfield  {author} {\bibinfo {author} {\bibfnamefont {N.}~\bibnamefont
  {Gnan}}\ and\ \bibinfo {author} {\bibfnamefont {E.}~\bibnamefont
  {Zaccarelli}},\ }\bibfield  {title} {\bibinfo {title} {{The microscopic role
  of deformation in the dynamics of soft colloids}},\ }\href
  {https://doi.org/10.1038/s41567-019-0480-1} {\bibfield  {journal} {\bibinfo
  {journal} {Nat. Phys.}\ }\textbf {\bibinfo {volume} {15}},\ \bibinfo {pages}
  {683} (\bibinfo {year} {2019})}\BibitemShut {NoStop}%
\bibitem [{\citenamefont {Poling-Skutvik}\ \emph {et~al.}(2019)\citenamefont
  {Poling-Skutvik}, \citenamefont {Slim}, \citenamefont {Narayanan},
  \citenamefont {Conrad},\ and\ \citenamefont
  {Krishnamoorti}}]{Poling-Skutvik2019}%
  \BibitemOpen
  \bibfield  {author} {\bibinfo {author} {\bibfnamefont {R.}~\bibnamefont
  {Poling-Skutvik}}, \bibinfo {author} {\bibfnamefont {A.~H.}\ \bibnamefont
  {Slim}}, \bibinfo {author} {\bibfnamefont {S.}~\bibnamefont {Narayanan}},
  \bibinfo {author} {\bibfnamefont {J.~C.}\ \bibnamefont {Conrad}},\ and\
  \bibinfo {author} {\bibfnamefont {R.}~\bibnamefont {Krishnamoorti}},\
  }\bibfield  {title} {\bibinfo {title} {{Soft Interactions Modify the
  Diffusive Dynamics of Polymer-Grafted Nanoparticles in Solutions of Free
  Polymer}},\ }\href {https://doi.org/10.1021/acsmacrolett.9b00294} {\bibfield
  {journal} {\bibinfo  {journal} {ACS Macro Lett.}\ }\textbf {\bibinfo {volume}
  {8}},\ \bibinfo {pages} {917} (\bibinfo {year} {2019})}\BibitemShut {NoStop}%
\bibitem [{\citenamefont {Holmqvist}\ and\ \citenamefont
  {N{\"{a}}gele}(2010)}]{Holmqvist2010}%
  \BibitemOpen
  \bibfield  {author} {\bibinfo {author} {\bibfnamefont {P.}~\bibnamefont
  {Holmqvist}}\ and\ \bibinfo {author} {\bibfnamefont {G.}~\bibnamefont
  {N{\"{a}}gele}},\ }\bibfield  {title} {\bibinfo {title} {{Long-Time Dynamics
  of Concentrated Charge-Stabilized Colloids}},\ }\href
  {https://doi.org/10.1103/PhysRevLett.104.058301} {\bibfield  {journal}
  {\bibinfo  {journal} {Phys. Rev. Lett.}\ }\textbf {\bibinfo {volume} {104}},\
  \bibinfo {pages} {058301} (\bibinfo {year} {2010})}\BibitemShut {NoStop}%
\bibitem [{\citenamefont {Senff}\ and\ \citenamefont
  {Richtering}(1999)}]{Senff1999}%
  \BibitemOpen
  \bibfield  {author} {\bibinfo {author} {\bibfnamefont {H.}~\bibnamefont
  {Senff}}\ and\ \bibinfo {author} {\bibfnamefont {W.}~\bibnamefont
  {Richtering}},\ }\bibfield  {title} {\bibinfo {title} {{Temperature sensitive
  microgel suspensions: Colloidal phase behavior and rheology of soft
  spheres}},\ }\href {https://doi.org/10.1063/1.479430} {\bibfield  {journal}
  {\bibinfo  {journal} {J. Chem. Phys.}\ }\textbf {\bibinfo {volume} {111}},\
  \bibinfo {pages} {1705} (\bibinfo {year} {1999})}\BibitemShut {NoStop}%
\bibitem [{\citenamefont {Smith}\ \emph {et~al.}(2006)\citenamefont {Smith},
  \citenamefont {Alexeev}, \citenamefont {Verberg},\ and\ \citenamefont
  {Balazs}}]{Smith2006}%
  \BibitemOpen
  \bibfield  {author} {\bibinfo {author} {\bibfnamefont {K.~A.}\ \bibnamefont
  {Smith}}, \bibinfo {author} {\bibfnamefont {A.}~\bibnamefont {Alexeev}},
  \bibinfo {author} {\bibfnamefont {R.}~\bibnamefont {Verberg}},\ and\ \bibinfo
  {author} {\bibfnamefont {A.~C.}\ \bibnamefont {Balazs}},\ }\bibfield  {title}
  {\bibinfo {title} {{Designing a simple ratcheting system to sort
  microcapsules by mechanical properties}},\ }\href
  {https://doi.org/10.1021/la0610093} {\bibfield  {journal} {\bibinfo
  {journal} {Langmuir}\ }\textbf {\bibinfo {volume} {22}},\ \bibinfo {pages}
  {6739} (\bibinfo {year} {2006})}\BibitemShut {NoStop}%
\bibitem [{\citenamefont {Finken}\ and\ \citenamefont
  {Seifert}(2006)}]{Finken2006}%
  \BibitemOpen
  \bibfield  {author} {\bibinfo {author} {\bibfnamefont {R.}~\bibnamefont
  {Finken}}\ and\ \bibinfo {author} {\bibfnamefont {U.}~\bibnamefont
  {Seifert}},\ }\bibfield  {title} {\bibinfo {title} {{Wrinkling of
  microcapsules in shear flow}},\ }\href
  {https://doi.org/10.1088/0953-8984/18/15/L04} {\bibfield  {journal} {\bibinfo
   {journal} {J. Phys. Condens. Matter}\ }\textbf {\bibinfo {volume} {18}},\
  \bibinfo {pages} {185} (\bibinfo {year} {2006})},\ \Eprint
  {https://arxiv.org/abs/0601589} {0601589 [cond-mat]} \BibitemShut {NoStop}%
\bibitem [{\citenamefont {Gross}\ \emph {et~al.}(2014)\citenamefont {Gross},
  \citenamefont {Kr{\"{u}}ger},\ and\ \citenamefont {Varnik}}]{Gross2014}%
  \BibitemOpen
  \bibfield  {author} {\bibinfo {author} {\bibfnamefont {M.}~\bibnamefont
  {Gross}}, \bibinfo {author} {\bibfnamefont {T.}~\bibnamefont
  {Kr{\"{u}}ger}},\ and\ \bibinfo {author} {\bibfnamefont {F.}~\bibnamefont
  {Varnik}},\ }\bibfield  {title} {\bibinfo {title} {{Rheology of dense
  suspensions of elastic capsules: Normal stresses, yield stress, jamming and
  confinement effects}},\ }\href {https://doi.org/10.1039/c4sm00081a}
  {\bibfield  {journal} {\bibinfo  {journal} {Soft Matter}\ }\textbf {\bibinfo
  {volume} {10}},\ \bibinfo {pages} {4360} (\bibinfo {year} {2014})},\ \Eprint
  {https://arxiv.org/abs/1401.2914} {1401.2914} \BibitemShut {NoStop}%
\bibitem [{\citenamefont {Gao}\ \emph {et~al.}(2012)\citenamefont {Gao},
  \citenamefont {Hu},\ and\ \citenamefont {Casta{\~{n}}eda}}]{Gao2012}%
  \BibitemOpen
  \bibfield  {author} {\bibinfo {author} {\bibfnamefont {T.}~\bibnamefont
  {Gao}}, \bibinfo {author} {\bibfnamefont {H.~H.}\ \bibnamefont {Hu}},\ and\
  \bibinfo {author} {\bibfnamefont {P.~P.}\ \bibnamefont {Casta{\~{n}}eda}},\
  }\bibfield  {title} {\bibinfo {title} {{Shape Dynamics and Rheology of Soft
  Elastic Particles in a Shear Flow}},\ }\href
  {https://doi.org/10.1103/PhysRevLett.108.058302} {\bibfield  {journal}
  {\bibinfo  {journal} {Phys. Rev. Lett.}\ }\textbf {\bibinfo {volume} {108}},\
  \bibinfo {pages} {058302} (\bibinfo {year} {2012})}\BibitemShut {NoStop}%
\bibitem [{\citenamefont {Danker}\ \emph {et~al.}(2009)\citenamefont {Danker},
  \citenamefont {Vlahovska},\ and\ \citenamefont {Misbah}}]{Danker2009}%
  \BibitemOpen
  \bibfield  {author} {\bibinfo {author} {\bibfnamefont {G.}~\bibnamefont
  {Danker}}, \bibinfo {author} {\bibfnamefont {P.~M.}\ \bibnamefont
  {Vlahovska}},\ and\ \bibinfo {author} {\bibfnamefont {C.}~\bibnamefont
  {Misbah}},\ }\bibfield  {title} {\bibinfo {title} {{Vesicles in Poiseuille
  Flow}},\ }\href {https://doi.org/10.1103/PhysRevLett.102.148102} {\bibfield
  {journal} {\bibinfo  {journal} {Phys. Rev. Lett.}\ }\textbf {\bibinfo
  {volume} {102}},\ \bibinfo {pages} {148102} (\bibinfo {year} {2009})},\
  \Eprint {https://arxiv.org/abs/0809.4028} {0809.4028} \BibitemShut {NoStop}%
\bibitem [{\citenamefont {Abreu}\ \emph {et~al.}(2014)\citenamefont {Abreu},
  \citenamefont {Levant}, \citenamefont {Steinberg},\ and\ \citenamefont
  {Seifert}}]{Abreu2014}%
  \BibitemOpen
  \bibfield  {author} {\bibinfo {author} {\bibfnamefont {D.}~\bibnamefont
  {Abreu}}, \bibinfo {author} {\bibfnamefont {M.}~\bibnamefont {Levant}},
  \bibinfo {author} {\bibfnamefont {V.}~\bibnamefont {Steinberg}},\ and\
  \bibinfo {author} {\bibfnamefont {U.}~\bibnamefont {Seifert}},\ }\bibfield
  {title} {\bibinfo {title} {{Fluid vesicles in flow}},\ }\href
  {https://doi.org/10.1016/j.cis.2014.02.004} {\bibfield  {journal} {\bibinfo
  {journal} {Adv. Colloid Interface Sci.}\ }\textbf {\bibinfo {volume} {208}},\
  \bibinfo {pages} {129} (\bibinfo {year} {2014})},\ \Eprint
  {https://arxiv.org/abs/1311.7341} {1311.7341} \BibitemShut {NoStop}%
\bibitem [{\citenamefont {Calabrese}\ \emph {et~al.}(2020)\citenamefont
  {Calabrese}, \citenamefont {{Da Silva}}, \citenamefont {Schmitt},
  \citenamefont {Hossain}, \citenamefont {Scott},\ and\ \citenamefont
  {Edler}}]{Calabrese2020a}%
  \BibitemOpen
  \bibfield  {author} {\bibinfo {author} {\bibfnamefont {V.}~\bibnamefont
  {Calabrese}}, \bibinfo {author} {\bibfnamefont {M.~A.}\ \bibnamefont {{Da
  Silva}}}, \bibinfo {author} {\bibfnamefont {J.}~\bibnamefont {Schmitt}},
  \bibinfo {author} {\bibfnamefont {K.~M.}\ \bibnamefont {Hossain}}, \bibinfo
  {author} {\bibfnamefont {J.~L.}\ \bibnamefont {Scott}},\ and\ \bibinfo
  {author} {\bibfnamefont {K.~J.}\ \bibnamefont {Edler}},\ }\bibfield  {title}
  {\bibinfo {title} {{Charge-driven interfacial gelation of cellulose
  nanofibrils across the water/oil interface}},\ }\href
  {https://doi.org/10.1039/c9sm01551e} {\bibfield  {journal} {\bibinfo
  {journal} {Soft Matter}\ }\textbf {\bibinfo {volume} {16}},\ \bibinfo {pages}
  {357} (\bibinfo {year} {2020})}\BibitemShut {NoStop}%
\bibitem [{\citenamefont {Bergstr{\"{o}}m}(1996)}]{Bergstrom1996}%
  \BibitemOpen
  \bibfield  {author} {\bibinfo {author} {\bibfnamefont {L.}~\bibnamefont
  {Bergstr{\"{o}}m}},\ }\bibfield  {title} {\bibinfo {title} {{Rheological
  properties of Al2O3-SiC whisker composite suspensions}},\ }\href
  {https://doi.org/10.1007/BF00355933} {\bibfield  {journal} {\bibinfo
  {journal} {J. Mater. Sci.}\ }\textbf {\bibinfo {volume} {31}},\ \bibinfo
  {pages} {5257} (\bibinfo {year} {1996})}\BibitemShut {NoStop}%
\bibitem [{\citenamefont {Simha}(1940)}]{Simha1940}%
  \BibitemOpen
  \bibfield  {author} {\bibinfo {author} {\bibfnamefont {R.}~\bibnamefont
  {Simha}},\ }\bibfield  {title} {\bibinfo {title} {{The Influence of Brownian
  Movement on the Viscosity of Solutions.}},\ }\href
  {https://doi.org/10.1021/j150397a004} {\bibfield  {journal} {\bibinfo
  {journal} {J. Phys. Chem.}\ }\textbf {\bibinfo {volume} {44}},\ \bibinfo
  {pages} {25} (\bibinfo {year} {1940})}\BibitemShut {NoStop}%
\bibitem [{\citenamefont {Palanisamy}\ and\ \citenamefont {den
  Otter}(2019)}]{Palanisamy2019}%
  \BibitemOpen
  \bibfield  {author} {\bibinfo {author} {\bibfnamefont {D.}~\bibnamefont
  {Palanisamy}}\ and\ \bibinfo {author} {\bibfnamefont {W.~K.}\ \bibnamefont
  {den Otter}},\ }\bibfield  {title} {\bibinfo {title} {{Intrinsic viscosities
  of non-spherical colloids by Brownian dynamics simulations}},\ }\href
  {https://doi.org/10.1063/1.5127001} {\bibfield  {journal} {\bibinfo
  {journal} {J. Chem. Phys.}\ }\textbf {\bibinfo {volume} {151}},\ \bibinfo
  {pages} {184902} (\bibinfo {year} {2019})}\BibitemShut {NoStop}%
\bibitem [{\citenamefont {Shewan}\ and\ \citenamefont
  {Stokes}(2014)}]{Shewan2014}%
  \BibitemOpen
  \bibfield  {author} {\bibinfo {author} {\bibfnamefont {H.~M.}\ \bibnamefont
  {Shewan}}\ and\ \bibinfo {author} {\bibfnamefont {J.~R.}\ \bibnamefont
  {Stokes}},\ }\bibfield  {title} {\bibinfo {title} {{Analytically predicting
  the viscosity of hard sphere suspensions from the particle size
  distribution}},\ }\href {https://doi.org/10.1016/j.jnnfm.2014.09.002}
  {\bibfield  {journal} {\bibinfo  {journal} {J. Nonnewton. Fluid Mech.}\
  }\textbf {\bibinfo {volume} {222}},\ \bibinfo {pages} {72} (\bibinfo {year}
  {2014})}\BibitemShut {NoStop}%
\bibitem [{\citenamefont {Woutersen}\ and\ \citenamefont {{De
  Kruif}}(1991)}]{Woutersen1991}%
  \BibitemOpen
  \bibfield  {author} {\bibinfo {author} {\bibfnamefont {A.~T.}\ \bibnamefont
  {Woutersen}}\ and\ \bibinfo {author} {\bibfnamefont {C.~G.}\ \bibnamefont
  {{De Kruif}}},\ }\bibfield  {title} {\bibinfo {title} {{The rheology of
  adhesive hard sphere dispersions}},\ }\href
  {https://doi.org/10.1063/1.460734} {\bibfield  {journal} {\bibinfo  {journal}
  {J. Chem. Phys.}\ }\textbf {\bibinfo {volume} {94}},\ \bibinfo {pages} {5739}
  (\bibinfo {year} {1991})}\BibitemShut {NoStop}%
\bibitem [{\citenamefont {Lu}\ \emph {et~al.}(2006)\citenamefont {Lu},
  \citenamefont {Conrad}, \citenamefont {Wyss}, \citenamefont {Schofield},\
  and\ \citenamefont {Weitz}}]{Lu2006}%
  \BibitemOpen
  \bibfield  {author} {\bibinfo {author} {\bibfnamefont {P.~J.}\ \bibnamefont
  {Lu}}, \bibinfo {author} {\bibfnamefont {J.~C.}\ \bibnamefont {Conrad}},
  \bibinfo {author} {\bibfnamefont {H.~M.}\ \bibnamefont {Wyss}}, \bibinfo
  {author} {\bibfnamefont {A.~B.}\ \bibnamefont {Schofield}},\ and\ \bibinfo
  {author} {\bibfnamefont {D.~A.}\ \bibnamefont {Weitz}},\ }\bibfield  {title}
  {\bibinfo {title} {{Fluids of Clusters in Attractive Colloids}},\ }\href
  {https://doi.org/10.1103/PhysRevLett.96.028306} {\bibfield  {journal}
  {\bibinfo  {journal} {Phys. Rev. Lett.}\ }\textbf {\bibinfo {volume} {96}},\
  \bibinfo {pages} {028306} (\bibinfo {year} {2006})}\BibitemShut {NoStop}%
\bibitem [{\citenamefont {Lu}\ \emph {et~al.}(2008)\citenamefont {Lu},
  \citenamefont {Zaccarelli}, \citenamefont {Ciulla}, \citenamefont
  {Schofield}, \citenamefont {Sciortino},\ and\ \citenamefont
  {Weitz}}]{Lu2008}%
  \BibitemOpen
  \bibfield  {author} {\bibinfo {author} {\bibfnamefont {P.~J.}\ \bibnamefont
  {Lu}}, \bibinfo {author} {\bibfnamefont {E.}~\bibnamefont {Zaccarelli}},
  \bibinfo {author} {\bibfnamefont {F.}~\bibnamefont {Ciulla}}, \bibinfo
  {author} {\bibfnamefont {A.~B.}\ \bibnamefont {Schofield}}, \bibinfo {author}
  {\bibfnamefont {F.}~\bibnamefont {Sciortino}},\ and\ \bibinfo {author}
  {\bibfnamefont {D.~A.}\ \bibnamefont {Weitz}},\ }\bibfield  {title} {\bibinfo
  {title} {{Gelation of particles with short-range attraction}},\ }\href
  {https://doi.org/10.1038/nature06931} {\bibfield  {journal} {\bibinfo
  {journal} {Nature}\ }\textbf {\bibinfo {volume} {453}},\ \bibinfo {pages}
  {499} (\bibinfo {year} {2008})}\BibitemShut {NoStop}%
\bibitem [{\citenamefont {Sprakel}\ \emph {et~al.}(2011)\citenamefont
  {Sprakel}, \citenamefont {Lindstr{\"{o}}m}, \citenamefont {Kodger},\ and\
  \citenamefont {Weitz}}]{Sprakel2011}%
  \BibitemOpen
  \bibfield  {author} {\bibinfo {author} {\bibfnamefont {J.}~\bibnamefont
  {Sprakel}}, \bibinfo {author} {\bibfnamefont {S.~B.}\ \bibnamefont
  {Lindstr{\"{o}}m}}, \bibinfo {author} {\bibfnamefont {T.~E.}\ \bibnamefont
  {Kodger}},\ and\ \bibinfo {author} {\bibfnamefont {D.~A.}\ \bibnamefont
  {Weitz}},\ }\bibfield  {title} {\bibinfo {title} {{Stress Enhancement in the
  Delayed Yielding of Colloidal Gels}},\ }\href
  {https://doi.org/10.1103/PhysRevLett.106.248303} {\bibfield  {journal}
  {\bibinfo  {journal} {Phys. Rev. Lett.}\ }\textbf {\bibinfo {volume} {106}},\
  \bibinfo {pages} {248303} (\bibinfo {year} {2011})}\BibitemShut {NoStop}%
\bibitem [{\citenamefont {Divoux}\ \emph {et~al.}(2013)\citenamefont {Divoux},
  \citenamefont {Grenard},\ and\ \citenamefont {Manneville}}]{Divoux2013}%
  \BibitemOpen
  \bibfield  {author} {\bibinfo {author} {\bibfnamefont {T.}~\bibnamefont
  {Divoux}}, \bibinfo {author} {\bibfnamefont {V.}~\bibnamefont {Grenard}},\
  and\ \bibinfo {author} {\bibfnamefont {S.}~\bibnamefont {Manneville}},\
  }\bibfield  {title} {\bibinfo {title} {{Rheological hysteresis in soft glassy
  materials}},\ }\href {https://doi.org/10.1103/PhysRevLett.110.018304}
  {\bibfield  {journal} {\bibinfo  {journal} {Phys. Rev. Lett.}\ }\textbf
  {\bibinfo {volume} {110}},\ \bibinfo {pages} {018304} (\bibinfo {year}
  {2013})},\ \Eprint {https://arxiv.org/abs/1207.3953} {1207.3953} \BibitemShut
  {NoStop}%
\bibitem [{\citenamefont {Radhakrishnan}\ \emph {et~al.}(2017)\citenamefont
  {Radhakrishnan}, \citenamefont {Divoux}, \citenamefont {Manneville},\ and\
  \citenamefont {Fielding}}]{Radhakrishnan2017}%
  \BibitemOpen
  \bibfield  {author} {\bibinfo {author} {\bibfnamefont {R.}~\bibnamefont
  {Radhakrishnan}}, \bibinfo {author} {\bibfnamefont {T.}~\bibnamefont
  {Divoux}}, \bibinfo {author} {\bibfnamefont {S.}~\bibnamefont {Manneville}},\
  and\ \bibinfo {author} {\bibfnamefont {S.~M.}\ \bibnamefont {Fielding}},\
  }\bibfield  {title} {\bibinfo {title} {{Understanding rheological hysteresis
  in soft glassy materials}},\ }\href {https://doi.org/10.1039/c6sm02581a}
  {\bibfield  {journal} {\bibinfo  {journal} {Soft Matter}\ }\textbf {\bibinfo
  {volume} {13}},\ \bibinfo {pages} {1834} (\bibinfo {year} {2017})},\ \Eprint
  {https://arxiv.org/abs/1611.03148} {1611.03148} \BibitemShut {NoStop}%
\bibitem [{\citenamefont {Boromand}\ \emph {et~al.}(2017)\citenamefont
  {Boromand}, \citenamefont {Jamali},\ and\ \citenamefont
  {Maia}}]{Boromand2017}%
  \BibitemOpen
  \bibfield  {author} {\bibinfo {author} {\bibfnamefont {A.}~\bibnamefont
  {Boromand}}, \bibinfo {author} {\bibfnamefont {S.}~\bibnamefont {Jamali}},\
  and\ \bibinfo {author} {\bibfnamefont {J.~M.}\ \bibnamefont {Maia}},\
  }\bibfield  {title} {\bibinfo {title} {{Structural fingerprints of yielding
  mechanisms in attractive colloidal gels}},\ }\href
  {https://doi.org/10.1039/C6SM00750C} {\bibfield  {journal} {\bibinfo
  {journal} {Soft Matter}\ }\textbf {\bibinfo {volume} {13}},\ \bibinfo {pages}
  {458} (\bibinfo {year} {2017})}\BibitemShut {NoStop}%
\bibitem [{\citenamefont {Koumakis}\ and\ \citenamefont
  {Petekidis}(2011)}]{Koumakis2011}%
  \BibitemOpen
  \bibfield  {author} {\bibinfo {author} {\bibfnamefont {N.}~\bibnamefont
  {Koumakis}}\ and\ \bibinfo {author} {\bibfnamefont {G.}~\bibnamefont
  {Petekidis}},\ }\bibfield  {title} {\bibinfo {title} {{Two Step Yielding in
  Attractive Colloids: Transition from Gels to Attractive Glasses}},\ }\href
  {https://doi.org/10.1039/C0SM00957A} {\bibfield  {journal} {\bibinfo
  {journal} {Soft Matter}\ }\textbf {\bibinfo {volume} {7}},\ \bibinfo {pages}
  {2456} (\bibinfo {year} {2011})}\BibitemShut {NoStop}%
\bibitem [{\citenamefont {Studart}\ \emph {et~al.}(2011)\citenamefont
  {Studart}, \citenamefont {Amstad},\ and\ \citenamefont
  {Gauckler}}]{Studart2011}%
  \BibitemOpen
  \bibfield  {author} {\bibinfo {author} {\bibfnamefont {A.~R.}\ \bibnamefont
  {Studart}}, \bibinfo {author} {\bibfnamefont {E.}~\bibnamefont {Amstad}},\
  and\ \bibinfo {author} {\bibfnamefont {L.~J.}\ \bibnamefont {Gauckler}},\
  }\bibfield  {title} {\bibinfo {title} {{Yielding of weakly attractive
  nanoparticle networks}},\ }\href {https://doi.org/10.1039/c1sm05598d}
  {\bibfield  {journal} {\bibinfo  {journal} {Soft Matter}\ }\textbf {\bibinfo
  {volume} {7}},\ \bibinfo {pages} {6408} (\bibinfo {year} {2011})}\BibitemShut
  {NoStop}%
\bibitem [{\citenamefont {Roy}\ and\ \citenamefont
  {Tirumkudulu}(2020)}]{Roy2020}%
  \BibitemOpen
  \bibfield  {author} {\bibinfo {author} {\bibfnamefont {S.}~\bibnamefont
  {Roy}}\ and\ \bibinfo {author} {\bibfnamefont {M.~S.}\ \bibnamefont
  {Tirumkudulu}},\ }\bibfield  {title} {\bibinfo {title} {{Micro-mechanical
  theory of shear yield stress for strongly flocculated colloidal gel}},\
  }\href {https://doi.org/10.1039/c9sm01784d} {\bibfield  {journal} {\bibinfo
  {journal} {Soft Matter}\ }\textbf {\bibinfo {volume} {2}},\ \bibinfo {pages}
  {1801} (\bibinfo {year} {2020})}\BibitemShut {NoStop}%
\bibitem [{\citenamefont {{Saez Cabezas}}\ \emph {et~al.}(2018)\citenamefont
  {{Saez Cabezas}}, \citenamefont {Ong}, \citenamefont {Jadrich}, \citenamefont
  {Lindquist}, \citenamefont {Agrawal}, \citenamefont {Truskett},\ and\
  \citenamefont {Milliron}}]{SaezCabezas2018}%
  \BibitemOpen
  \bibfield  {author} {\bibinfo {author} {\bibfnamefont {C.~A.}\ \bibnamefont
  {{Saez Cabezas}}}, \bibinfo {author} {\bibfnamefont {G.~K.}\ \bibnamefont
  {Ong}}, \bibinfo {author} {\bibfnamefont {R.~B.}\ \bibnamefont {Jadrich}},
  \bibinfo {author} {\bibfnamefont {B.~A.}\ \bibnamefont {Lindquist}}, \bibinfo
  {author} {\bibfnamefont {A.}~\bibnamefont {Agrawal}}, \bibinfo {author}
  {\bibfnamefont {T.~M.}\ \bibnamefont {Truskett}},\ and\ \bibinfo {author}
  {\bibfnamefont {D.~J.}\ \bibnamefont {Milliron}},\ }\bibfield  {title}
  {\bibinfo {title} {{Gelation of plasmonic metal oxide nanocrystals by
  polymer-induced depletion attractions}},\ }\href
  {https://doi.org/10.1073/pnas.1806927115} {\bibfield  {journal} {\bibinfo
  {journal} {Proc. Natl. Acad. Sci. U.S.A.}\ }\textbf {\bibinfo {volume}
  {115}},\ \bibinfo {pages} {8925} (\bibinfo {year} {2018})}\BibitemShut
  {NoStop}%
\bibitem [{\citenamefont {Bergenholtz}\ \emph {et~al.}(2003)\citenamefont
  {Bergenholtz}, \citenamefont {Poon},\ and\ \citenamefont
  {Fuchs}}]{Bergenholtz2003}%
  \BibitemOpen
  \bibfield  {author} {\bibinfo {author} {\bibfnamefont {J.}~\bibnamefont
  {Bergenholtz}}, \bibinfo {author} {\bibfnamefont {W.~C.}\ \bibnamefont
  {Poon}},\ and\ \bibinfo {author} {\bibfnamefont {M.}~\bibnamefont {Fuchs}},\
  }\bibfield  {title} {\bibinfo {title} {{Gelation in model colloid-polymer
  mixtures}},\ }\href {https://doi.org/10.1021/la0340089} {\bibfield  {journal}
  {\bibinfo  {journal} {Langmuir}\ }\textbf {\bibinfo {volume} {19}},\ \bibinfo
  {pages} {4493} (\bibinfo {year} {2003})}\BibitemShut {NoStop}%
\bibitem [{\citenamefont {Yu}\ and\ \citenamefont
  {Somasundaran}(1996)}]{Yu1996}%
  \BibitemOpen
  \bibfield  {author} {\bibinfo {author} {\bibfnamefont {X.}~\bibnamefont
  {Yu}}\ and\ \bibinfo {author} {\bibfnamefont {P.}~\bibnamefont
  {Somasundaran}},\ }\bibfield  {title} {\bibinfo {title} {{Role of polymer
  conformation in interparticle-bridging dominated flocculation}},\ }\href
  {https://doi.org/10.1006/jcis.1996.0033} {\bibfield  {journal} {\bibinfo
  {journal} {J. Colloid Interface Sci.}\ }\textbf {\bibinfo {volume} {177}},\
  \bibinfo {pages} {283} (\bibinfo {year} {1996})}\BibitemShut {NoStop}%
\bibitem [{\citenamefont {Biggs}(1995)}]{Biggs1995}%
  \BibitemOpen
  \bibfield  {author} {\bibinfo {author} {\bibfnamefont {S.}~\bibnamefont
  {Biggs}},\ }\bibfield  {title} {\bibinfo {title} {{Steric and Bridging Forces
  between Surfaces Bearing Adsorbed Polymer: An Atomic Force Microscopy
  Study}},\ }\href {https://doi.org/10.1021/la00001a028} {\bibfield  {journal}
  {\bibinfo  {journal} {Langmuir}\ }\textbf {\bibinfo {volume} {11}},\ \bibinfo
  {pages} {156} (\bibinfo {year} {1995})}\BibitemShut {NoStop}%
\bibitem [{\citenamefont {Daly}\ \emph {et~al.}(2020)\citenamefont {Daly},
  \citenamefont {Riley}, \citenamefont {Segura},\ and\ \citenamefont
  {Burdick}}]{Daly2020}%
  \BibitemOpen
  \bibfield  {author} {\bibinfo {author} {\bibfnamefont {A.~C.}\ \bibnamefont
  {Daly}}, \bibinfo {author} {\bibfnamefont {L.}~\bibnamefont {Riley}},
  \bibinfo {author} {\bibfnamefont {T.}~\bibnamefont {Segura}},\ and\ \bibinfo
  {author} {\bibfnamefont {J.~A.}\ \bibnamefont {Burdick}},\ }\bibfield
  {title} {\bibinfo {title} {{Hydrogel microparticles for biomedical
  applications}},\ }\href {https://doi.org/10.1038/s41578-019-0148-6}
  {\bibfield  {journal} {\bibinfo  {journal} {Nat. Rev. Mater.}\ }\textbf
  {\bibinfo {volume} {5}},\ \bibinfo {pages} {20} (\bibinfo {year}
  {2020})}\BibitemShut {NoStop}%
\end{thebibliography}%

\end{document}